\documentclass[11pt]{article}

\usepackage[a4paper,margin=1in]{geometry}
\usepackage{amsmath,amssymb,bm}
\usepackage{graphicx}
\usepackage{booktabs}
\usepackage{array}
\usepackage{tabularx}
\usepackage{xcolor}
\usepackage{subcaption}
\usepackage{hyperref}
\usepackage{url}
\usepackage{placeins}

\graphicspath{{figures/}}
\hypersetup{
  colorlinks=true,
  linkcolor=blue!45!black,
  citecolor=blue!45!black,
  urlcolor=blue!55!black
}

\newcommand{\pfus}{P_{\rm fus}}
\newcommand{\pheat}{P_{\rm heat}}
\newcommand{\ptrans}{P_{\rm trans}}
\newcommand{\pbrem}{P_{\rm brem}}
\newcommand{\pcycl}{P_{\rm cycl}}
\newcommand{\pline}{P_{\rm line}}
\newcommand{\pn}{P_{\rm n}}
\newcommand{\qfus}{Q_{\rm fus}}
\newcommand{\tauE}{\tau_E}
\newcommand{\Vp}{V_{\rm p}}
\newcommand{\Sp}{S_{\rm p}}
\newcommand{\Sw}{S_{\rm w}}
\newcommand{\Zeff}{Z_{\rm eff}}
\newcommand{\gbr}{g_{\rm Br}}
\newcommand{\dd}{\,{\rm d}}
\newcommand{\sv}{\langle\sigma v\rangle}
\newcommand{\rhoVol}{\rho_{\rm vol}}
\newcommand{\Rmaj}{R_0}
\newcommand{\amin}{a}
\title{VSC: A Zero-Dimensional Fusion Design Platform for Multiple Magnetic Configurations}
\author{
Zekun Wang$^{1}$, Huasheng Xie$^{1,*}$, Feng Zhang$^{1}$,
Jian Bao$^{1}$, and Ming Yang$^{1}$\\[0.6em]
\normalsize $^{1}$Beijing VeloAlpha Technology Co., Ltd., Beijing, China\\[-0.1em]
\small $^*$Corresponding author: \href{mailto:huashengxie@gmail.com}{huashengxie@gmail.com}
}
\date{\today}

\begin{document}
\maketitle

\begin{abstract}
The VeloAlpha System Code (VSC) is a computational framework for
zero-dimensional fusion power-balance studies across five magnetic-confinement
configurations: tokamaks, magnetic mirrors, field-reversed configurations
(FRCs), dipoles, and stellarators. A common power-balance formulation connects
fusion production, charged-particle deposition, radiation, transport loss,
external heating, and fusion gain, while each configuration retains its own
geometry, profile weights, confinement model, and operating constraints. The
same solver interface supports both single-point calculations and
two-dimensional plasma operating contour (POPCON) scans, producing fusion and heating powers, gain,
radiation and transport losses, geometry quantities, and
configuration-specific validity indicators. VSC therefore makes it possible
to study how assumptions about density, temperature, magnetic field,
confinement, and geometry shape the accessible operating space of different
fusion concepts within one traceable framework. By combining reduced-order
physics models with a unified computational platform, VSC enables rapid
assessment and comparative analysis of candidate fusion reactor concepts
during the early design stage.
\end{abstract}

\section{Introduction}

A fusion system code combines plasma-physics assumptions, device geometry,
power-balance relations, and operating constraints into a self-consistent
description of a candidate fusion device. It plays an important role between
conceptual physics and detailed engineering design: rapid parameter-space
exploration helps establish the required device size, magnetic field, density,
temperature, confinement, and auxiliary heating, while identifying promising
operating windows and the assumptions that require higher-fidelity analysis.
In a zero-dimensional system code,
spatially varying quantities are represented through volume-integrated power
balances together with prescribed profile and geometry factors. Such a model
cannot replace equilibrium, transport, stability, or engineering analysis;
instead, it provides a fast and transparent foundation for early device design
before resource-intensive integrated calculations are undertaken
\cite{xie2023ignition}.

Plasma operating contour (POPCON) diagrams are a natural output of this approach because they display
power balance and operating constraints over two scanned variables. ENN's
published proton--boron roadmap illustrates how system-code studies can
connect spherical-torus physics to an experiment--ignition--power development
sequence; its subsequent response clarifies the density convention, profile
assumptions, and updated power-balance modeling used in that work
\cite{liu2024ennroadmap,xie2025response}. Modern tokamak tools such as cfspopcon and
OpenPOPCON make POPCON calculations reproducible and inspectable
\cite{cfspopcon,openpopcon}. The Fusion Synthesis Engine (FUSE) provides a
broader integrated environment for plasma, engineering, control, and pilot-
plant design \cite{meneghini2024fuse}. PROCESS links plasma assumptions to a much
broader plant model and includes dedicated stellarator power-plant models in
addition to its tokamak workflow \cite{processcode}.

Outside tokamaks, the computational landscape is more topology-specific.
MCTrans++ provides zero-dimensional modeling for centrifugal mirrors, while
recent tandem-mirror studies combine reduced confinement models with
device-specific performance predictions \cite{mctranscode,frank2025hammir}.
FRC research uses dedicated equilibrium and stability tools, including
two-dimensional equilibrium reconstruction \cite{ma2021}. Dipole studies are
still commonly organized around analytic flux-shell, interchange-stability,
and reactor-scaling models \cite{hasegawa1987}. Beyond the systems-level
stellarator model in PROCESS, stellarator workflows place substantial
emphasis on three-dimensional equilibrium and magnetic-field
optimization through tools such as VMEC, near-axis quasisymmetry methods, and
DESC \cite{hirshman1983,garren1991,landreman2019,panici2022}. These tools provide
substantial configuration-specific depth, but they address different parts of
the modeling chain. A shared zero-dimensional power-balance study spanning all
five magnetic topologies considered here remains less common.

The rapid growth of commercial fusion programs has placed many proposed
devices at an early design stage, where assumptions must be explored before
resource-intensive integrated calculations are justified. A transparent tool
for preliminary device studies is therefore timely. VSC addresses this need
with one solver interface for tokamaks, magnetic
mirrors, FRCs, dipoles, and stellarators. It combines a shared top-level power
balance with configuration-specific geometry, profiles, confinement models,
and operating limits. The goal is not to rank these concepts by one number,
but to expose how each assumed operating point is constructed. For every
valid point, VSC reports fusion power, required external heating, fusion gain,
radiation and transport losses, geometry factors, operating-window status,
and diagnostics specific to the selected configuration. Single-point
calculations and two-dimensional scans use the same underlying solver path.

V1 of VSC is available online at
\url{https://hub.veloalpha.cn/vsc/}. This initial version provides the core
functions used here: parameter input, single-point calculation, geometry
inspection, and two-dimensional POPCON scans.

This work describes the common power balance, the configuration-dependent
models, the numerical scan workflow, and the available verification evidence.
It also presents one ITER-named operating-space comparison with FUSE.
That comparison is limited to qualitative contour morphology and
operating-space trends; it is not a pointwise numerical benchmark because
the two calculations use different density coordinates, power definitions,
and scenario assumptions. Broader comparisons require matched
geometry, grids, density conventions, impurities, alpha deposition,
confinement scalings, radiation models, and operating-window definitions.

The remainder of this work is organized as follows. Section~\ref{sec:arch}
introduces the VSC framework, numerical interfaces, common zero-dimensional
power balance, profiles, coordinates, and geometry weights.
Section~\ref{sec:models} describes the tokamak, magnetic-mirror, FRC, dipole,
and stellarator models and identifies the physics retained by each branch.
Section~\ref{sec:popcon} presents the POPCON workflow, operating-window logic,
browser interface, and representative runtime. Section~\ref{sec:verification}
then summarizes the verification evidence, the FUSE ITER comparison, and the
limitations and trust boundary of the calculations. The final section draws
the main results and presents the outlook toward higher-dimensional fusion-
device design.

\section{Computational Workflow and Numerical Methods}
\label{sec:arch}

\begin{figure}[t]
\centering
\includegraphics[width=\linewidth]{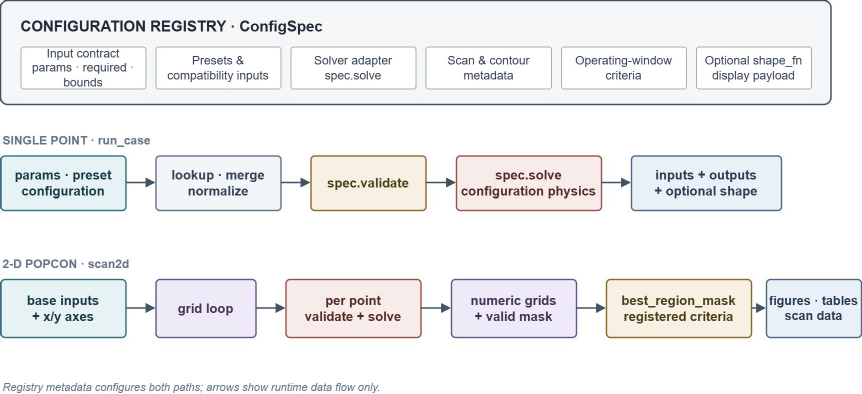}
\caption{Computation sequence used by all five magnetic configurations. Each
configuration model supplies its physical input domain, geometry and profile
weights, power-balance calculation, scan variables, displayed contours, and
operating limits. \texttt{run\_case} selects the model, normalizes and validates the input,
and one \texttt{spec.solve} call. \texttt{scan2d} independently applies the
same \texttt{spec.validate}/\texttt{spec.solve} kernel at every grid point,
collects finite numeric outputs, and constructs a validity mask before
\texttt{best\_region\_mask} applies the stated operating criteria. Geometry and
profile weights remain inside each configuration solver rather than in the
scan engine.}
\label{fig:architecture}
\end{figure}

Each magnetic configuration is described by a self-contained physics model
that specifies its admissible inputs, geometry, profile integrals,
confinement relations, operating limits, and default scan variables. The
single-point calculation and two-dimensional scan evaluate the same physical
model, through the internal routines \texttt{run\_case} and \texttt{scan2d},
respectively. Figure~\ref{fig:architecture}
summarizes the data flow shared by application interfaces, batch scans, and
reproducible figure generation.

The narrow interface has three consequences. First, adding or modifying one
configuration does not require a new POPCON loop. Second, diagnostic names
can remain configuration-specific while still appearing in a common scan
object. Third, geometry, profile weights, and scalar outputs are traceable:
publication figures can be regenerated from source-generated numerical outputs
and the same public interfaces used by the application.

The core single-point contract is
\begin{equation}
  {\cal S}_c(\bm{x}) \rightarrow
  \{\pfus,\pheat,\qfus,\pbrem,\pcycl,\pline,\ptrans,
  \Vp,\Sw,\bm{r}_c\},
\end{equation}
where \(c\) denotes the selected magnetic configuration, \(\bm{x}\) is an
input-domain-checked vector, and \(\bm{r}_c\) denotes
configuration-specific redline diagnostics. A scan over two inputs \(x_i\)
and \(x_j\) is the independent
grid evaluation
\begin{equation}
  G_{mn} =
  \begin{cases}
    {\cal S}_c(\bm{x}_{mn}), & \bm{x}_{mn}\in{\cal D}_c,\\
    {\rm invalid}, & \bm{x}_{mn}\notin{\cal D}_c,
  \end{cases}
\end{equation}
where \({\cal D}_c\) is the input and solver domain for configuration
\(c\). Invalid points are carried through the plotting layer as masked
values. A default operating window is then a predicate on finite outputs,
\begin{equation}
  {\cal W}_{mn} =
  \left[
  \bigwedge_k r_k(G_{mn}) \ge r_{k,\min}
  \right]
  \wedge
  \left[
  \bigwedge_l r_l(G_{mn}) \le r_{l,\max}
  \right],
\end{equation}
with the lower and upper thresholds specified for that configuration.
This formulation is intentionally modest: a mask marks a candidate operating window
under declared assumptions, not a feasible plant design.

\subsection{Common Zero-Dimensional Power Accounting}
\label{sec:power}

This section is best read as a power-accounting overview rather than as a derivation of
new confinement physics. Equations~\eqref{eq:pfus}--\eqref{eq:pheat} define
where fusion power enters, how charged particles and neutrons are separated,
and how required heating is formed after radiation, transport, and deposited
self-heating are combined. The branch-specific sections then specify the
geometry weights, field moments, and confinement closures that feed this
shared account. Table~\ref{tab:symbol-guide} lists the symbols most likely to
be confused when comparing POPCON panels.

\begin{table}[t]
\centering
\caption{Minimal symbol guide for reading the power-account section.}
\label{tab:symbol-guide}
\small
\begin{tabularx}{\linewidth}{@{}p{0.18\linewidth}p{0.33\linewidth}X@{}}
\toprule
Symbol & Meaning & Reading note \\
\midrule
\(\pfus\) & fusion power & Source term before charged-particle deposition and neutron escape are included in the power account. \\
\(\pheat\) & required external heating power & Negative or zero values mark an ignition-like power-balance state, not a plant-level operating claim. \\
\(\qfus\) & \(\pfus/\pheat\) & Displayed gain is conditional on all supplied closures and masks. \\
\(\gbr\) & total bremsstrahlung Gaunt factor & Species-resolved in the current model; it is not a scalar-\(\Zeff\) correction. \\
\(\Zeff\) & effective ion charge & Reported for diagnostics and older-formula comparison; the Xie bremsstrahlung kernel uses the full ion list. \\
\(\tauE\) & energy confinement time & Input in manual mode; solved only where a branch implements a predictive or semi-predictive closure. \\
\(\Vp,\Sp,\Sw\) & plasma volume, plasma surface, wall surface & Defined by each branch's native geometry, not by a single common radius. \\
\(G_{B25}\) or \(c_{B25}\) & \(B^{2.5}\) radiation moment & Branch-specific; cross-configuration cyclotron comparisons require caution. \\
\({\cal W}\) & operating-window mask & Conditional operating criterion, not a feasibility prediction. \\
\bottomrule
\end{tabularx}
\end{table}
 
All configurations share the same accounting skeleton. A fuel model selects
reaction charges, mass numbers, a representative fast charged product for
two-temperature diagnostics, the charged-particle energy fraction, the
neutron fraction, a like-reactant double-counting flag, and the released
energy. Bosch-Hale reactivities are used for the principal thermonuclear
branches \cite{bosch1992}; the two implemented \(p\)--\(^{11}{\rm B}\)
branches expose the two implemented fits as separate reactivity-model choices.
Table~\ref{tab:reaction-account} records the reaction-accounting constants
used by the source. The principal light-ion reaction energies follow the NRL
Plasma Formulary, while the two proton--boron cross-section choices follow
Nevins--Swain and Sikora--Weller, respectively
\cite{nrl2019,nevins2000,sikora2016}.

\begin{table}[t]
\centering
\small
\caption{Reaction energy accounting used in VSC. \(f_{\rm ion}\)
is the fraction of released fusion energy carried by charged products and
therefore available for local deposition before configuration-specific loss
corrections.}
\label{tab:reaction-account}
\begin{tabularx}{\linewidth}{@{}c>{\raggedright\arraybackslash}Xcc>{\raggedright\arraybackslash}X@{}}
\toprule
\texttt{icase} & Reaction & \(E_{\rm fus}\) [MeV] & \(f_{\rm ion}\) & Note \\
\midrule
1 & D--T & 17.59 & 0.20 & 3.5 MeV alpha heating plus escaping neutron energy. \\
2 & D--D & \(\tfrac{1}{2}(3.27+4.04)\) & \(\simeq 0.66\) & Sum of the two DD branches with like-reactant correction. \\
3 & D--\(^{3}{\rm He}\) & 18.35 & 1.00 & All primary products charged. \\
4 & \(p\)--\(^{11}{\rm B}\) (Nevins) & 8.68 & 1.00 & Source-side Nevins-style reactivity option. \\
5 & \(p\)--\(^{11}{\rm B}\) (Sikora--Weller) & 8.68 & 1.00 & Interpolated Sikora--Weller cross-section data. \\
6 & Catalyzed D--D & \(\tfrac{1}{2}43.25\) & \(26.73/43.25\) & Aggregate of the DD daughters followed by complete D--T and D--$^3$He burn. \\
\bottomrule
\end{tabularx}
\end{table}

For a fuel mixture with impurity and helium dilution, the source first forms
the fuel-pair density \(n_{12,0}=f_{12}n_{i0}^{\rm tot}\), with
\(f_{12}=1-f_{\rm He}-f_{\rm imp}\). It then sets
\(x_1=1,x_2=1\) for like-reactant channels and
\(x_1=f_1,x_2=1-f_1\) otherwise, so \(n_{10}=x_1n_{12,0}\) and
\(n_{20}=x_2n_{12,0}\). The quasineutrality and dilution quantities are
\begin{align}
  n_{\rm He,0} &= f_{\rm He}n_{i0}^{\rm tot},\qquad
  n_{\rm imp,0}=f_{\rm imp}n_{i0}^{\rm tot},\\
  n_{e0} &= \frac{Z_1n_{10}+Z_2n_{20}}{1+\delta_{12}}
  +2n_{\rm He,0}+Z_{\rm imp}n_{\rm imp,0},\\
  \Zeff &=
  \frac{(Z_1^2n_{10}+Z_2^2n_{20})/(1+\delta_{12})
  +4n_{\rm He,0}+Z_{\rm imp}^2n_{\rm imp,0}}
  {n_{e0}},
  \label{eq:zeff}\\
  M &= \frac{x_1A_1+x_2A_2}{1+\delta_{12}},
\end{align}
where \(\delta_{12}=1\) for like-reactant channels and zero otherwise.
The mean mass number \(M\) enters the tokamak confinement scalings. The scalar
\(\Zeff\) is still reported for collisional diagnostics and comparison with
older reduced formulas, but the current bremsstrahlung model uses the resolved
ion list
\begin{equation}
  {\cal I}_0 =
  \left\{
  \left(\frac{n_{10}}{1+\delta_{12}},Z_1\right),
  \left(\frac{n_{20}}{1+\delta_{12}},Z_2\right),
  (n_{\rm He,0},2),(n_{\rm imp,0},Z_{\rm imp})
  \right\},
  \label{eq:ion-list}
\end{equation}
with zero-density entries omitted. For reactants with central densities
\(n_{10}\) and \(n_{20}\), the fusion power is written
\begin{equation}
  \pfus =
  \frac{E_{\rm fus}}{1+\delta_{12}}\,
  n_{10} n_{20}\, \Phi_{\rm fus}\, \Vp ,
  \label{eq:pfus}
\end{equation}
where \(E_{\rm fus}\) is the reaction energy and \(\Phi_{\rm fus}\) is the
profile-weighted reactivity integral. For the closed-torus profile family,
\begin{equation}
  n(\rho)=n_0(1-\rho^2)^{S_n}, \qquad
  T(\rho)=T_0(1-\rho^2)^{S_T},
\end{equation}
the source evaluates
\begin{equation}
  \Phi_{\rm fus}
  =
  2 f_\sigma
  \int_0^1
  (1-\rho^2)^{2S_n}
  \sv[T(\rho)]\,\rho\,\dd\rho .
\end{equation}
Like-particle branches include the usual factor avoiding double counting,
implemented in the reaction metadata.

The reaction energy is then split into charged-particle and neutron
channels,
\begin{equation}
  P_{\rm charged}=f_{\rm ion}\pfus,\qquad
  \pn=(1-f_{\rm ion})\pfus ,
  \label{eq:charged-split}
\end{equation}
where \(f_{\rm ion}\) is reaction dependent. In branches with loss-cone
physics, \(P_{\rm charged,dep}\) may be smaller than
\(P_{\rm charged}\).

The stored thermal energy is
\begin{equation}
  E_{\rm th} =
  \frac{3}{2}\, \Vp\,{\rm keV}
  \left[
  n_{i0}T_{i0} + n_{e0}T_{e0}
  \right] f_{nT},
  \qquad
  f_{nT} = \frac{1}{1+S_n+S_T}
\end{equation}
for the shared profile family, with analogous geometry-specific weights
used by FRC and dipole branches. Bremsstrahlung, line radiation, cyclotron
or synchrotron radiation, and transport losses are then combined as
\begin{equation}
  \pheat =
  \ptrans + \pbrem + \pcycl + \pline - P_{\rm charged,dep}.
  \label{eq:pheat}
\end{equation}
The fusion gain reported by the scan layer is
\begin{equation}
  \qfus = \frac{\pfus}{\pheat},
\end{equation}
with a raw value and an ignition flag retained when \(\pheat \le 0\).
The code therefore avoids interpreting a clipped display value as physical
evidence of a viable operating point.

The shared bremsstrahlung term is profile weighted and now follows the
species-resolved Xie thermal-average Gaunt-factor fit \cite{xie2024brems}.
For a local plasma state, the emissivity is
\begin{equation}
  p_{\rm Br}(\rho)=
  C_{\rm Xie}\,n_e(\rho)^2\sqrt{T_e(\rho)}\,
  \gbr(\rho),
  \label{eq:brem}
\end{equation}
where \(C_{\rm Xie}=4.86\times10^{-37}\,{\rm W\,m^3\,keV^{-1/2}}\),
\(T_e\) is in keV, and
\begin{equation}
  \gbr =
  \sum_{i\in{\cal I}}
  Z_i^2\,\frac{n_i}{n_e}\,g_{ei}\!\left(t,Z_i\right)
  +g_{ee}(t),
  \qquad
  t=\frac{T_e}{511}.
  \label{eq:xie-gaunt-mixture}
\end{equation}
For a single pure ion species this reduces to
\(\gbr=Zg_{ei}(t,Z)+g_{ee}(t)\). The calculation uses the unified fitting
formulation of Xie: \(g_{ei}=c_0(f_{\rm nr}-f_Z)+f_r\), where
\(c_0=2\sqrt{3}/\pi\) and the non-relativistic, relativistic, and high-\(Z\)
correction terms follow the coefficients in Ref.~\cite{xie2024brems}; the
electron-electron term is evaluated with the corresponding fitted
\(g_{ee}(t)\). The important modeling point is that the electron-ion Gaunt
factor depends on \(Z_i\), so replacing the mixture by a single \(\Zeff\)
would not be rigorous for multi-ion plasmas or impurity-bearing advanced-fuel
cases \cite{xie2024brems}.

The remaining configuration dependence enters through the geometry-specific
volume element:
\begin{equation}
  \pbrem=\int_{\Vp} p_{\rm Br}\,\dd V.
  \label{eq:brem-integral}
\end{equation}
For the shared toroidal profile family this is evaluated as
\(2\Vp\int_0^1 p_{\rm Br}(\rho)\rho\,\dd\rho\); FRC, dipole, mirror, and
imported-equilibrium branches use their native volume weights with the same
local emissivity. Optional impurity radiation uses the Mavrin
coronal-equilibrium total cooling coefficient \(L_z^{\rm tot}(T_e)\)
\cite{mavrin2018}. Because the impurity electron-ion bremsstrahlung share is
already counted in Eq.~\eqref{eq:xie-gaunt-mixture}, the added
line/recombination channel is netted as
\begin{equation}
  L_z^{\rm net}(T_e)=
  \max\!\left[
  L_z^{\rm tot}(T_e)
  - C_{\rm Xie}Z_{\rm imp}^2\sqrt{T_e}\,
  g_{ei}\!\left(T_e/511,Z_{\rm imp}\right),\,0
  \right],
  \label{eq:line-net}
\end{equation}
and then integrated over the profile:
\begin{equation}
  \pline =
  \Vp \int n_e(\rho)n_{\rm imp}(\rho)
  L_z^{\rm net}[T_e(\rho)]\,\frac{\dd V}{\Vp}.
\end{equation}
Cyclotron radiation is evaluated either from the branch-specific formula or
from an explicit cyclotron-loss time. In the browser interface this choice is
shown as \emph{use input tauC (off = configuration fast formula)}:
\begin{equation}
  \pcycl =
  \begin{cases}
  {\cal P}_{\rm cyc}(n_e,T_e,B,\Sw,\hbox{profile}), & {\rm formula\ mode},\\
  W_e/\tau_C, & {\rm prescribed}\ \tau_C\ {\rm mode}.
  \end{cases}
  \label{eq:tauc}
\end{equation}
This keeps a user-prescribed \(\tau_C\) used for sensitivity tests distinct
from the transport confinement time \(\tauE\). The nonuniform-field
correction enters through a volume moment of the magnetic-field strength. The
front-end control \emph{include magnetic-field nonuniformity in cyclotron
loss} enables this correction; when it is off, the corresponding branch uses
its uniform reference-field treatment.
In branches that use the shared Trubnikov/Albajar fast estimate, the base
scaling is
\begin{equation}
{\cal P}_{\rm cyc}=
4.14\times10^{-7}
n_{\rm eff}^{1/2}T_{\rm eff}^{2.5}B_0^{2.5}
(1-R_w)^{1/2}a_{\rm eff}^{-1/2}
\left(1+\frac{2.5T_{\rm eff}}{511}\right)
\Vp\,c_{B25},
\label{eq:cyclotron-base}
\end{equation}
with \(n_{\rm eff}\) in \(10^{20}{\rm m^{-3}}\), \(T_{\rm eff}\) in keV, and
the branch-specific \(c_{B25}\) replaced by a dimensional
\(\int B^{2.5}\dd V\) moment in the mirror branch. The convexity of
\(B^{2.5}\) matters: in the FRC branch \(G_{B25}=\langle |B/B_e|^{2.5}\rangle\)
is computed independently and is not equal to \(G_B^{2.5}\).
Tokamak scans can use the Miller \(B_T\propto1/R\) moment or, for imported
G-EQDSK equilibria, the actual \(|B(R,Z)|\) moment. Stellarator scans use the
near-axis \(B/B_0=1+\bar{\eta}a\rho\cos\theta\) moment or, for imported VMEC
equilibria, the equilibrium's magnetic-field-strength harmonics and Jacobian. Mirror,
FRC, and dipole branches keep their native axial, rigid-rotor, or shell
weights inside the radiation or \(W_e/\tau_C\) calculation
\cite{trubnikov1979,albajar2001}.

The two-temperature treatment is explicit. Each branch reports a Stix
critical energy \(E_c\), fast-product ion-deposition fraction
\(f_{\rm fast,ion}\),
ion-electron equilibration time \(\tau_{ei}\), and collisional exchange power
\begin{equation}
  P_{ei}=\frac{3}{2}\frac{n_i(T_i-T_e)\,{\rm keV}}{\tau_{ei}}\,\Vp,
\end{equation}
with configuration-specific profile weights. The Stix deposition calculation
therefore distinguishes ion-channel and electron-channel alpha heating, so self-consistent
electron-channel solves use the thermalized charged-particle part rather than
assigning all charged-product power to the bulk. Tokamak, mirror, and
stellarator branches expose an electron-channel self-consistency mode
(\(f_T=0\) or \(T_{e0}=0\), depending on the branch) in which the electron
temperature is found from
\begin{equation}
  (1-f_{\rm fast,ion})P_{\rm charged,dep}+P_{ei}
  +f_{{\rm aux},e}\max(\pheat,0)
  =
  \pbrem+\pcycl+\pline+E_{{\rm th},e}/\tauE .
\end{equation}
If no bracketed root is found, the source reports that the solution was pinned
to a bracket endpoint rather than presenting it as a converged
two-temperature solution
\cite{stix1972,nrl2019,wesson2011}.

The most consequential modeling choice is the meaning of \(\tauE\). The
browser control is labeled \emph{use input tauE (off = self-consistent
losses)}. In its default on state, \(\tauE\) is an input. The result should be
read as: if this plasma achieved the supplied confinement time, the 0-D
power account would close as shown. Some branches also expose predictive or
semi-predictive confinement closures. For tokamaks, setting this control to
off solves for the confinement time that satisfies
\(H_{\rm scl}(\tauE)=H_{\rm fac}\), where the selected scaling can be IPB98,
spherical-tokamak, or ITPA20 \cite{iter1999,verdoolaege2021}. The solve is a
one-dimensional bracket-and-bisect problem, with a monotone shortcut when
\(T_e\) is fixed and a scan bracket when the electron channel is nested.
For stellarators, the corresponding off state targets
\(H_{\rm ISS04}=H_{\rm fac}\), while Sudo density margins remain redline
diagnostics \cite{yamada2005,sudo1990}. For magnetic mirrors, disabling the
manual \(\tauE\) input switches transport loss to the assembled
Pastukhov/gas-dynamic/radial channel time. For FRCs, the LSX-style
confinement time is available in the source outputs. The dipole branch has no
predictive \(\tauE\) scaling in the present framework, so the input semantics
are explicit.

\subsection{Profiles, Coordinates, and Geometry Weights}

\begin{figure}[t]
\centering
\includegraphics[width=\linewidth]{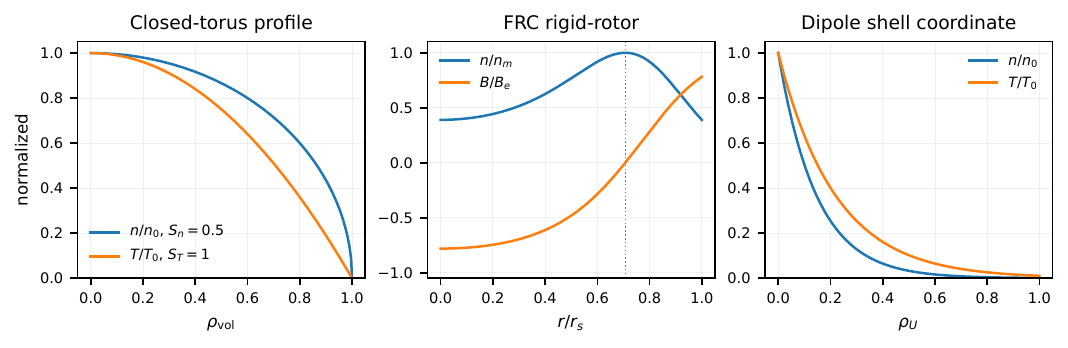}
\caption{Representative normalized profiles and their native coordinates.
Closed toroidal branches use a volume-radius coordinate so that
\((1-\rho^2)^S\) profiles have analytic volume averages. The FRC panel shows
the signed rigid-rotor magnetic field and density profile used in the
pressure-balance closure. Dipole profiles are monotone functions of the
flux-tube-volume coordinate rather than of a minor radius.}
\label{fig:profiles}
\end{figure}

The common interface does not force all configurations into the same radial
coordinate. Closed toroidal branches use a volume-radius coordinate,
\begin{equation}
  \rhoVol = \sqrt{\frac{V(\rho)}{\Vp}},
\end{equation}
which reduces to the usual minor-radius coordinate for self-similar circular
surfaces. FRCs use the rigid-rotor radial coordinate. Dipoles use a
flux-shell volume coordinate. Figure~\ref{fig:profiles} compares these
coordinate conventions.

This distinction is not cosmetic. In Eq.~\eqref{eq:pfus}, \(\Vp\) and
\(\Phi_{\rm fus}\) are only meaningful if the profile coordinate and volume
element match the magnetic topology. VSC therefore lets each
configuration own its geometry weights while sharing the top-level power
accounting.
The volume-radius convention also preserves a common analytic form for
profile averages across noncircular and three-dimensional surfaces, so
changes in geometry are not inadvertently absorbed into profile exponents.

\section{Geometric and Transport Models}
\label{sec:models}

Table~\ref{tab:config-models} summarizes the five magnetic-configuration
models evaluated through the same power-balance procedure. The following subsections describe the geometry
and closure logic used by each branch. The emphasis is on what is computed
and what the result should be trusted to mean.

Before the per-branch detail, Table~\ref{tab:config-maturity} states the
asymmetry in branch maturity directly: whether a predictive confinement closure
exists, what verification is available, what physics is omitted, and the
resulting assessment level. This asymmetry is the single most important
caveat when reading the combined POPCON panel in Sec.~\ref{sec:popcon}: the
tokamak branch carries predictive scalings and imported equilibria, whereas the
dipole branch is parametric only. The shared interface unifies the software
contract and the top-level power account, not the underlying confinement
physics.

\begin{table*}[t]
\centering
\caption{Configuration models available through the same VSC calculation interface. The last column lists assessment quantities, not design guarantees.}
\label{tab:config-models}
\footnotesize
\setlength{\tabcolsep}{3pt}
\begin{tabular}{@{}p{0.11\textwidth}p{0.20\textwidth}p{0.19\textwidth}p{0.12\textwidth}p{0.24\textwidth}@{}}
\toprule
Config. & Geometry model & Closure mode & Default scan & Assessment quantities \\
\midrule
Tokamak & legacy, Miller, Cerfon--Freidberg, or imported G-EQDSK & manual $\tau_E$ or IPB98/ST/ITPA20 closure & $T_{i0}\times n_{i0}$ & $P_{\rm fus}$, $Q_{\rm fus}$, Greenwald, $q_{95}$, $\beta_N$, $H_{98}$ \\
Mirror & central cell plus $\sin^2$ throat contraction & manual $\tau_E$ or parallel loss channels & $T_{i0}\times n_{i0}$ & $P_{\rm fus}$, $Q_{\rm fus}$, $\beta$, Pastukhov/gas-dynamic tags \\
FRC & superellipse or Ma-Xie separatrix; flux-contour display & manual $\tau_E$ or LSX-style scaling & $T_i\times B_e$ & $P_{\rm fus}$, $Q_{\rm fus}$, $\bar{s}$, trapped-flux diagnostics \\
Dipole & point dipole or finite current-loop shells & manual $\tau_E$ in the present model & $T_{i0}\times n_0$ & $P_{\rm fus}$, $Q_{\rm fus}$, $\beta_{\rm in}$, shell-integrated radiation \\
Stellarator & simple/R2 near-axis, boundary Fourier, or VMEC/DESC & manual $\tau_E$ or ISS04/Sudo diagnostics & $T_{i0}\times n_{i0}$ & $P_{\rm fus}$, $Q_{\rm fus}$, Sudo margin, $H_{\rm ISS04}$, $\beta$ \\
\bottomrule
\end{tabular}
\end{table*}
 \begin{table*}[t]
\centering
\caption{Maturity of the five configuration branches. The branches are
\emph{not} equally predictive: they differ in whether a predictive confinement
closure exists, in the kind of verification available, and in the physics that
is deliberately omitted. The final column is a qualitative assessment label,
not a calibration guarantee.}
\label{tab:config-maturity}
\footnotesize
\setlength{\tabcolsep}{3pt}
\begin{tabularx}{\textwidth}{@{}p{0.10\textwidth}p{0.15\textwidth}p{0.20\textwidth}p{0.24\textwidth}p{0.15\textwidth}@{}}
\toprule
Branch & Predictive closure & Verification available & Key omissions & Assessment level \\
\midrule
Tokamak &
Yes: IPB98, ST, ITPA20 $\tau_E$ scalings &
Published $\tau_E$/Greenwald anchors; multiple geometry models; imported G-EQDSK &
No pedestal/ELM, current drive, bootstrap, or divertor constraint &
Highest \\
Mirror &
Yes: Pastukhov / gas-dynamic / radial end-loss &
Independent channel recompute; regime criteria; WHAM-order end-loss time &
Optimistic solid-angle $\alpha$ loss; classical radial transport; no sheath physics &
Medium \\
Stellarator &
Yes: ISS04 (with Sudo margin) &
Near-axis geometry cross-checked to pyQSC at $10^{-13}$; imported VMEC/DESC &
No neoclassical transport, island/divertor physics, or $\alpha$ confinement scaling &
Medium \\
FRC &
Partial: single LSX-style empirical scaling &
Analytic equilibrium moments vs numeric; average-$\beta$ theorem &
No tilt-mode evolution, flux-tube end shortening, or rotational instability &
Limited \\
Dipole &
None: $\tau_E$ is a user input &
Analytic shell geometry vs numeric; marginal-profile identities &
No predictive $\tau_E$; finite-orbit-width effects; equatorial-shell cyclotron proxy &
Parametric only \\
\bottomrule
\end{tabularx}
\end{table*}
 
\subsection{Tokamak}

\begin{figure}[t]
\centering
\includegraphics[width=\linewidth]{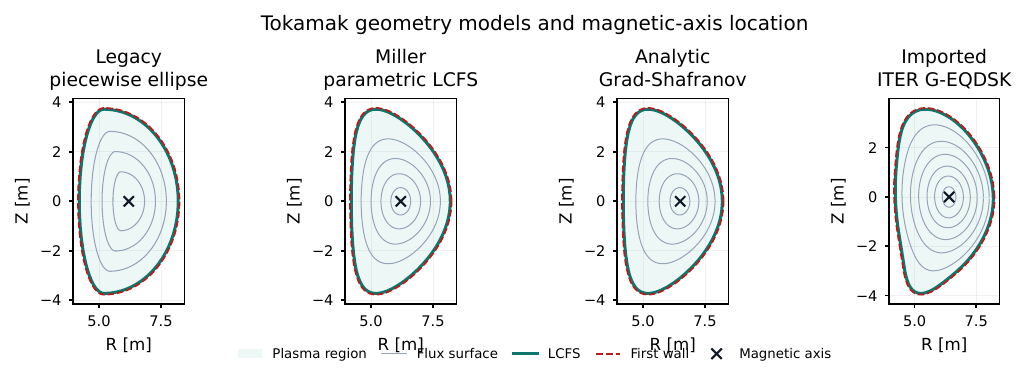}
\caption{Tokamak geometry models and magnetic-axis locations. The black
cross marks the axis reported by the source geometry payload; it is displaced
from the geometric center in the analytic Grad-Shafranov and imported
G-EQDSK cases. Pale cyan shading marks the plasma region inside the teal
last closed flux surface; red dashed curves indicate wall offsets.}
\label{fig:tokamak-geometry}
\end{figure}

The tokamak branch is the most complete POPCON-style branch in the present
framework. It exposes four geometry routes, shown in
Fig.~\ref{fig:tokamak-geometry}: a legacy piecewise double ellipse for
backward-compatible closed-form behavior, a Miller boundary with exact
surface-of-revolution metrics, an analytic Cerfon-Freidberg
Grad-Shafranov boundary, and imported G-EQDSK boundaries when an equilibrium
file is supplied \cite{cerfon2010}.

For a closed poloidal contour \((R(\theta),Z(\theta))\), the Miller and
equilibrium paths use the exact axisymmetric identities
\begin{align}
  \Vp &= \left| \pi \oint R^2\,\dd Z \right|,\\
  \Sp &= \oint 2\pi R\,\dd s,
  \qquad
  \dd s = \sqrt{(\dd R)^2+(\dd Z)^2}.
\end{align}
The Miller boundary is
\begin{align}
  R(\theta) &= \Rmaj + \amin
  \cos\left(\theta+\sin^{-1}\delta\,\sin\theta\right),\\
  Z(\theta) &= \kappa \amin \sin\theta ,
\end{align}
where \(\kappa\) is elongation and \(\delta\) is triangularity.
The unshifted elliptical limit used for quick interpretation is recovered by
\(\delta=0\),
\begin{equation}
  \frac{(R-\Rmaj)^2}{a^2}+\frac{Z^2}{(\kappa a)^2}=1,
  \qquad
  A=\frac{\Rmaj}{a}.
\end{equation}
This limit gives the leading closed-form geometric scales
\begin{equation}
  \Vp \simeq 2\pi^2 \kappa \Rmaj a^2,
  \qquad
  S_{\rm ell}\simeq 2\pi\Rmaj\,4\kappa a\,
  E\!\left(\sqrt{1-\kappa^{-2}}\right),
\end{equation}
for \(\kappa\ge1\), where \(E(\cdot)\) is the complete elliptic integral of
the second kind. These formulas are not used when a Miller, analytic
Grad-Shafranov, or imported equilibrium contour is available; they are included
to make the ellipse geometry behind the reduced tokamak limit explicit.
The legacy route is retained as the original closed-form compatibility fit,
rather than as a traced flux contour. With \(a=R_0/A\) and
\(A_d=R_0/(a+g)\), it uses
\begin{align}
  \Vp &=
  \left(2\pi^2\kappa(A-\delta)
  +\frac{16\pi}{3}\kappa\delta\right)a^3,\\
  \Sp &= \left(4\pi^2A\kappa^{0.65}-4\kappa\delta\right)a^2,\\
  \Sw &= \left(4\pi^2A_d\kappa^{0.65}-4\kappa\delta\right)(a+g)^2 .
\end{align}
The Miller and equilibrium routes are therefore the preferred paths when
shape factors, wall offsets, nested-volume radii, and nonuniform-field
moments are relevant.

The analytic Grad-Shafranov route solves for a Cerfon-Freidberg limiter
boundary. In normalized coordinates \(x=R/R_0\), \(y=Z/R_0\), the implemented
flux is
\begin{equation}
  \psi(x,y)=\psi_p(x)+\sum_{j=1}^{7}c_j\psi_j(x,y),
  \qquad
  \psi_p=\frac{x^4}{8}
  +A_{\rm CF}\left(\frac{x^2\ln x}{2}-\frac{x^4}{8}\right),
\end{equation}
where \(A_{\rm CF}=-0.155\) in the reference limiter and the homogeneous basis is
\begin{align}
\{\psi_j\}=\{&
1,\ x^2,\ y^2-x^2\ln x,\ x^4-4x^2y^2,\nonumber\\
&2y^4-9x^2y^2+3x^4\ln x-12x^2y^2\ln x,\nonumber\\
&x^6-12x^4y^2+8x^2y^4,\nonumber\\
&8y^6-140x^2y^4+75x^4y^2-15x^6\ln x
 +180x^4y^2\ln x-120x^2y^4\ln x\}.
\end{align}
The coefficients \(c_j\) are determined from \(\psi=0\) at the outer, inner,
and upper limiter points, a vanishing horizontal derivative at the upper point,
and the three corresponding curvature constraints. The LCFS is then the
\(\psi=0\) contour traced by ray bisection, and the branch reports the midplane
magnetic-axis displacement
\begin{equation}
  \Delta_{\rm Shaf}=R_{\rm axis}-R_0 .
\end{equation}
When a G-EQDSK file is imported, the LCFS, nested flux contours, axis,
boundary-derived volume, and optional real-field cyclotron correction are
read from the equilibrium object rather than reconstructed from \(R_0\),
\(a\), \(\kappa\), and \(\delta\). The imported normalized flux and field
moment use
\begin{align}
  \psi_N &= \frac{\psi(R,Z)-\psi_{\rm axis}}
  {\psi_{\rm LCFS}-\psi_{\rm axis}},\\
  B_R &= -\frac{1}{R}\frac{\partial\psi}{\partial Z},\qquad
  B_Z = \frac{1}{R}\frac{\partial\psi}{\partial R},\qquad
  B_\phi = \frac{F(\psi)}{R},\\
  {\cal G}_{B,p} &=
  \frac{\int_{\rm LCFS} (|B|/B_{\rm axis})^p\,2\pi R\,\dd R\,\dd Z}
       {\int_{\rm LCFS}2\pi R\,\dd R\,\dd Z},
  \qquad |B|=\sqrt{B_R^2+B_Z^2+B_\phi^2}.
\end{align}

The tokamak power account follows Eq.~\eqref{eq:pheat}. Radiation includes
species-resolved Xie bremsstrahlung, optional Mavrin-style impurity line
cooling without double counting the electron-ion bremsstrahlung contribution
\cite{xie2024brems,mavrin2018}, and an optional nonuniform-field cyclotron
correction. The branch reports
Greenwald ratio, cylindrical \(q\), shaped-edge \(q_{95}\), toroidal and
normalized beta, L-H threshold ratio, and confinement-quality factors.
Representative formulas include
\begin{align}
  \frac{\bar n}{n_{\rm Gw}} &=
  \frac{\bar n}{10^{20} I_p/(\pi \amin^2)},\\
  \beta_N &= 100\,\frac{\beta_T}{I_p/(\amin B_T)},\\
  q &= \frac{5B_T\amin^2\kappa}{R_0I_p},\\
  q_{95} &=
  \frac{5a^2B_T}{R_0I_p}
  \frac{1+\kappa^2(1+2\delta^2-1.2\delta^3)}{2}
  \frac{1.17-0.65\epsilon}{(1-\epsilon^2)^2},\\
  P_{\rm LH} &=
  0.0488
  \left(\frac{\bar n}{10^{20}}\right)^{0.717}
  B_T^{0.803}\Sp^{0.941}\left(\frac{2}{M}\right),
\end{align}
where \(I_p\) is in MA in the implemented convention, \(M\) is the mean
fuel mass number, and \(\epsilon=a/R_0\). The Greenwald density and L--H
threshold expressions follow Refs.~\cite{greenwald1988,martin2008}; the
confinement scalings below follow the ITER Physics Basis and the updated ITPA
database \cite{iter1999,verdoolaege2021}. When \(\tauE\) is supplied directly, \(H_{98}\),
\(H_{\rm ST}\), and \(H_{\rm ITPA20}\) are diagnostics. When the predictive
sentinel is used, the solver instead searches for a \(\tauE\) that gives the
requested confinement-quality factor.
The shaped-edge safety factor is evaluated separately from the cylindrical
reference \(q\); the code reports both because the kink-oriented \(q_{95}\)
redline and the historical cylindrical diagnostic are not interchangeable.

The three tokamak confinement scalings live side by side in the result:
\begin{align}
\tau_{98} &=
\frac{0.145 I_p^{0.93}R_0^{1.39}a^{0.58}\kappa^{0.78}
(\bar n/10^{20})^{0.41}B_T^{0.15}M^{0.19}}
{P_L^{0.69}},\\
\tau_{\rm ST} &=
0.066 I_p^{0.53}B_T^{1.05}
(\bar n/10^{19})^{0.65}R_0^{2.66}\kappa^{0.78}P_L^{-0.58},\\
\tau_{\rm ITPA20} &=
\frac{0.067M^{0.3}B_T^{-0.13}I_p^{1.29}R_0^{1.19}
(1+\delta)^{0.56}\kappa^{0.67}(\bar n/10^{19})^{0.15}}
{P_L^{0.644}},
\end{align}
where \(P_L=f_{\rm ion}\pfus+\pheat\) in the source convention. The predictive
mode solves \(H_{\rm scl}=\tauE/\tau_{\rm scl}=H_{\rm fac}\), with
\({\rm scl}\in\{98,{\rm ST},{\rm ITPA20}\}\), over the bracket
\(\tauE\in[10^{-3},50]\,{\rm s}\). If \(P_L\le0\), the scaling diagnostics are
zero or NaN rather than extrapolated through a singular power law. The
electron-temperature sentinel \(f_T=0\) can be nested inside this \(\tauE\)
root; the inner solve brackets the electron-channel residual and returns a
pinned flag when no sign-changing bracket exists. The branch also reports
an effective collisionality diagnostic,
\begin{equation}
  \nu_{\rm eff}=0.1\,\Zeff\, n_{19}R_0/\langle T_e\rangle^2,
\end{equation}
used for density-peaking context, but it is not a transport solver.

\subsection{Magnetic Mirror}

\begin{figure}[t]
\centering
\includegraphics[width=\linewidth]{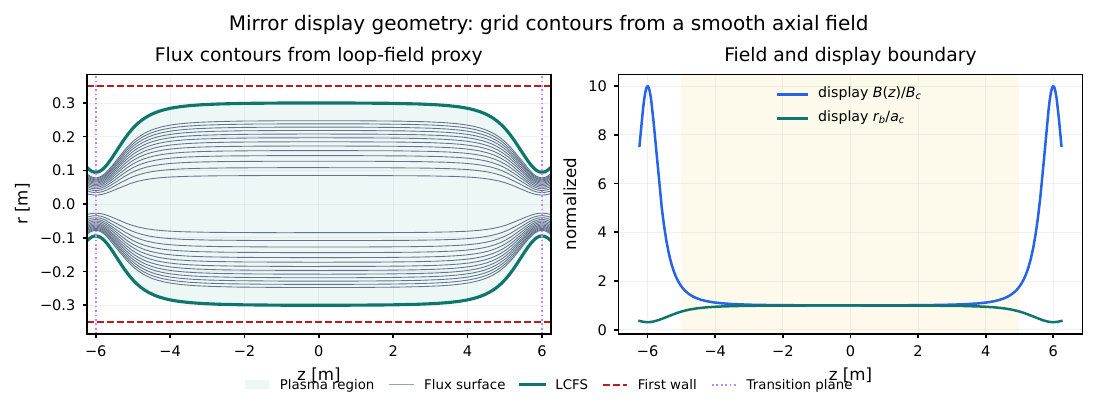}
\caption{Mirror display geometry and source field scale. The left panel
shows flux contours sampled from the loop-field-based display proxy, with the
teal contour marking the display boundary. The right panel shows the
corresponding normalized display field and boundary radius. The 0-D source
uses the analytic \(\sin^2\) throat volume and axial field moment.}
\label{fig:mirror-geometry}
\end{figure}

The mirror branch models a central cell plus a \(\sin^2\) throat
contraction in the 0-D power account. The display geometry in
Fig.~\ref{fig:mirror-geometry} follows the application convention: a smooth
axial-field proxy is constructed from a symmetric pair of loop fields with a
paraxial radial correction, and its flux function is sampled on a \((z,r)\)
grid. The plotted curves are contour levels of that grid. Scalar volume,
area, and cyclotron moments remain the analytic source quantities below. The
axial field ratio used by the 0-D source is
\begin{equation}
  \frac{B(z)}{B_c}
  =
  1+(R_m-1)
  \sin^2\left(\frac{\pi u}{2}\right),
  \qquad
  u = \frac{|z|-L_c/2}{L_{\rm th}},
\end{equation}
inside each throat, with \(u\) clipped to \([0,1]\). Flux conservation gives
\begin{equation}
  a(z) = a_c\sqrt{\frac{B_c}{B(z)}}.
\end{equation}
The end-region volume then integrates analytically, so
\begin{equation}
  \Vp = \pi a_c^2 L_c
  + \frac{2\pi a_c^2 L_{\rm th}}{\sqrt{R_{\rm mc}}},
\end{equation}
where \(R_{\rm mc}\) is the diamagnetically corrected mirror ratio used by
the solver.

The peak beta is computed from the vacuum field and peak pressure,
\begin{equation}
  \beta = \frac{2\mu_0 p_0}{B_{\rm vac}^2},
  \qquad
  B_c = B_{\rm vac}\sqrt{1-\beta},
  \qquad
  R_{\rm mc} = \frac{R_m}{\sqrt{1-\beta}}.
\end{equation}
The parallel confinement path combines Pastukhov, gas-dynamic, and radial
channels \cite{pastukhov1974}. With the default input-\(\tauE\) setting,
\(\ptrans=E_{\rm th}/\tauE\). In the loss-channel mode, the effective
particle time is assembled from the channel times. The ambipolar potential
model first estimates the ion barrier as
\begin{equation}
  \phi_i = T_{e0}\ln R_{\rm mirror},
\end{equation}
and then solves the electron barrier from \(y e^y=K\), where
\begin{equation}
  y=\frac{\phi_e}{T_{e0}},\qquad
  K=\sqrt{\frac{m_i}{m_e}}
  \left(\frac{T_{i0}}{T_{e0}}\right)^{3/2}
  \frac{\phi_i}{T_{i0}}
  \exp\!\left(\frac{\phi_i}{T_{i0}}\right).
\end{equation}
This represents the standard ambipolar picture: fast electrons would leave
first, so a positive plasma potential grows until electron and ion end losses
are compatible. With \(r=\phi_i/T_{i0}\), \(s=\sqrt{1+1/R_{\rm mc}}\), and
\(G=s\ln[(s+1)/(s-1)]\), the source Pastukhov channel is
\begin{equation}
\tau_{\rm Past}=
\frac{\sqrt{\pi}}{2}\tau_{ii}r e^rG
\left[
1+\frac{T_{i0}}{2\phi_i}
-\left(\frac{T_{i0}}{2\phi_i}\right)^2
\right]^{-1},
\end{equation}
inside its strong-barrier domain; a flagged fallback is used when the barrier
is too weak. The gas-dynamic and radial channels are
\begin{equation}
  \tau_{\rm gd}=\sqrt{\pi}R_{\rm mc}\frac{L_c}{v_{\rm th}}e^r,
  \qquad
  \tau_{\rho}=\left(\frac{a_c}{\rho_i}\right)^2\tau_{ii}.
\end{equation}
They combine as
\begin{equation}
  \tau_m^{-1}
  =
  \left(\tau_{\rm Past}+\tau_{\rm gd}\right)^{-1}
  + \tau_{\rho}^{-1},
\end{equation}
and the end-loss transport power includes both potential and thermal
energy per escaping particle. The branch also reports the throat power flux
and collector-diluted flux, because open-field end loading can dominate
concept evaluation even when the 0-D gain appears favorable.
The throat area used for this engineering diagnostic is
\begin{equation}
  A_{\rm th}=\frac{\pi a_c^2\sqrt{1-\beta}}{R_m}.
\end{equation}
The code also exposes the collisionality ratio
\begin{equation}
  \chi_{\rm coll}=\frac{\lambda_{ii}}{R_{\rm mc}L_c},
\end{equation}
which separates gas-dynamic-like and Pastukhov-like end-loss regimes. In
the self-consistent mirror path, the escaping-particle energy includes the
ambipolar potentials,
\begin{equation}
  \ptrans =
  \left[
  \frac{n_{i0}\phi_i+n_{e0}\phi_e}{1+S_n}
  +\frac{n_{i0}T_{i0}+n_{e0}T_{e0}}{1+S_n+S_T}
  \right]\frac{{\rm keV}\,\Vp}{\tau_m}.
\end{equation}
The self-consistent mirror mode also reduces charged-product self-heating by
the alpha deposition estimate
\begin{equation}
f_{\alpha,{\rm used}}=
\begin{cases}
1, & {\rm prescribed}\ \tauE\ {\rm mode},\\
\sqrt{1-1/R_{\rm mc}}, & {\rm loss\ cone\ estimate},\\
f_\alpha, & {\rm user\ supplied}.
\end{cases}
\end{equation}
The loss-cone value is an optimistic solid-angle estimate; it does not model
Coulomb scattering of slowing-down products into the loss cone.
For sensitivity studies we therefore treat it as an upper-deposition case.
A conservative comparison case uses \(0.5\sqrt{1-1/R_{\rm mc}}\). At
\(R_{\rm mc}=35\), for example, the optimistic estimate is \(0.986\), while
the conservative case is \(0.493\). Since the deposited charged-product term
enters Eq.~\eqref{eq:pheat} linearly, any mirror POPCON window that depends on
alpha self-heating should be read as robust only if it survives this reduction
or an explicitly supplied \(f_\alpha\) scan.
The mirror cyclotron model keeps the axial field variation inside the
radiation integral,
\begin{equation}
  {\cal M}_{B}=
  B_c^{2.5}\pi a_c^2L_c
  +2\int_{\rm throat}B(z)^{2.5}\pi a(z)^2\,\dd z ,
\end{equation}
so strong-field throat regions are not collapsed into a single central-field
factor.

\subsection{Field-Reversed Configuration}

\begin{figure}[t]
\centering
\includegraphics[width=\linewidth]{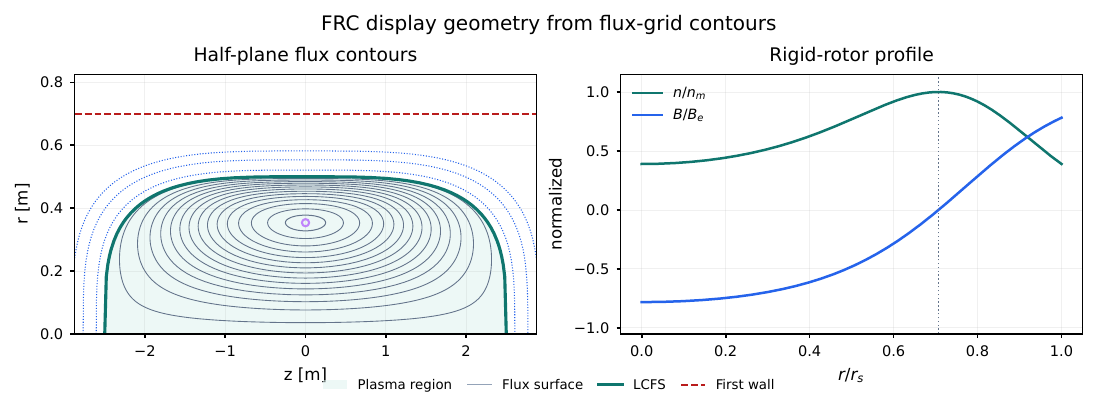}
\caption{FRC half-plane display geometry and rigid-rotor profile regenerated
from the source and front-end formulas. The left panel shows flux-like grid
contours inside the separatrix together with exterior SOL contours; the teal
curve is the separatrix used for the finite-length geometry family. The right
panel shows the rigid-rotor density and reversed-field profile used in the
0-D account.}
\label{fig:frc-geometry}
\end{figure}

The FRC branch is organized around the rigid-rotor equilibrium and a
finite-length separatrix family. Figure~\ref{fig:frc-geometry} shows the
half-plane convention used in this manuscript and by the application by
default: the separatrix and exterior scrape-off-layer (SOL) curves are contour
levels of smooth flux-like grid functions. An optional full-plane browser view
only mirrors this same half-plane data; it introduces no independent geometry.
These display contours do not replace the finite-length volume weights used by
the 0-D account.
The radial coordinate is \(x=r/r_s\), and the source implements
\begin{equation}
  u = 2x^2-1,\qquad
  \frac{n(x)}{n_m} = {\rm sech}^2(Ku),\qquad
  \frac{B(x)}{B_e} = \tanh(Ku).
  \label{eq:frc-rr}
\end{equation}
The parameter \(K\) is not a free peaking input. It is fixed by the
average-beta theorem,
\begin{equation}
  \frac{\tanh K}{K} = 1-\frac{x_s^2}{2},
  \qquad x_s=\frac{r_s}{r_w}.
\end{equation}
At the field null, pressure balance fixes the peak pressure rather than
letting density be an independent input:
\begin{equation}
  p_m=\frac{B_e^2}{2\mu_0},\qquad
  n_{i,m}=\frac{p_m}{(T_i+\zeta T_e)\,{\rm keV}},
  \qquad n_{e,m}=\zeta n_{i,m}.
\end{equation}
Thus a density-like user input in the FRC branch is a derived or diagnostic
quantity once \(B_e\), \(T_i\), and \(T_e\) are fixed; it is not an
independent density control. The result payload therefore identifies the FRC
density as the output of the rigid-rotor pressure-balance closure, so an
interface need not present it as an independent control.
For the cylindrical weighting limit, the profile factors used in the power
account are
\begin{align}
  G_1 &= \frac{\langle n\rangle}{n_m} = \frac{\tanh K}{K},\\
  G_2 &= \frac{\langle n^2\rangle}{n_m^2}
  = \frac{\tanh K-\tanh^3 K/3}{K},\\
  G_B &= \frac{\langle |B|\rangle}{B_e}
  = \frac{\ln\cosh K}{K},\\
  G_{B25} &= \left\langle |B/B_e|^{2.5}\right\rangle .
\end{align}
The last moment is evaluated numerically because \(G_{B25}\ne G_B^{2.5}\).
For the default FRC example in the source tests, \(G_B^{2.5}\) would
underestimate the synchrotron moment by roughly 47\%, a direct consequence
of Jensen's inequality for the convex function \(x^{2.5}\).

The separatrix volume is
\begin{equation}
  \left|\frac{2z}{l_s}\right|^p
  +\left|\frac{r}{r_s}\right|^p = 1,\qquad
  \Vp = C(p)\,\pi r_s^2 l_s,
  \qquad
  C(p)=\frac{1}{p}B\!\left(\frac{1}{p},1+\frac{2}{p}\right),
\end{equation}
where \(C(2)=2/3\) gives an ellipsoid of revolution and \(C(p)\to1\) gives a
racetrack-like finite-length limit. The optional Ma--Xie separatrix family
uses
\begin{equation}
  r(z)=r_s\left(1-\left|\frac{2z}{l_s}\right|^m\right)^{1/2},
  \qquad f_{\rm shape}=\frac{m}{m+1},
\end{equation}
which shares the same ellipse and racetrack limits \cite{ma2021}. The trapped
poloidal flux and diffusion comparison are reported as
\begin{equation}
  \phi_p=\frac{\pi r_s^2B_e}{2K}\ln\!\cosh K,\qquad
  \tau_\eta=\frac{\mu_0 r_s^2}{\eta},\qquad
  \frac{\tau_N}{\tau_\eta}.
\end{equation}
The branch also reports the kinetic parameter \(\bar{s}\), \(s/E\)
tilt-stability proxy, resistive and Bohm brackets, and the ratio between
confinement and flux-diffusion times. These are reduced-model diagnostics
motivated by FRC literature, not full kinetic-MHD stability calculations
\cite{tuszewski1988,steinhauer2011,ma2021}.
In predictive mode the LSX particle-confinement scaling is used as an
energy-confinement proxy after radiation is accounted for separately:
\begin{equation}
  \tauE =
  3.2\times10^{-15}
  E^{1/2}x_s^{0.8}r_s^{2.1}n_{e,m}^{0.6},
  \qquad E=\frac{l_s}{2r_s}.
\end{equation}
Here $x_s=r_s/r_w$. Hoffman and Slough reported this expression as an
empirical particle-lifetime scaling from the Large s Experiment
\cite{hoffman1993lsx}. VSC uses that particle lifetime as an
energy-confinement proxy after accounting for radiation separately; this is a
modeling approximation rather than a first-principles transport law.
The kinetic parameter and diffusion brackets are reported as
\begin{align}
  s&=\frac{r_s}{\rho_{ie}},\qquad s/E\lesssim3\hbox{--}4
  \quad{\rm as\ a\ tilt\ proxy},\\
  \tau_{\rm classical}&=\frac{r_s^2}{4D_{\rm cl}},\qquad
  D_{\rm cl}=\frac{2\eta_{\rm Sp}(n_iT_i+n_eT_e){\rm keV}}{B_e^2},\\
  \tau_{\rm Bohm}&=\frac{r_s^2}{4D_B},\qquad
  D_B=\frac{T_e[{\rm eV}]}{16B_e}.
\end{align}
The LSX-style \(\tauE\) proxy should be read as an empirical FRC estimate that can
be compared with these brackets, not as a first-principles transport closure.

\subsection{Dipole}

\begin{figure}[t]
\centering
\includegraphics[width=\linewidth]{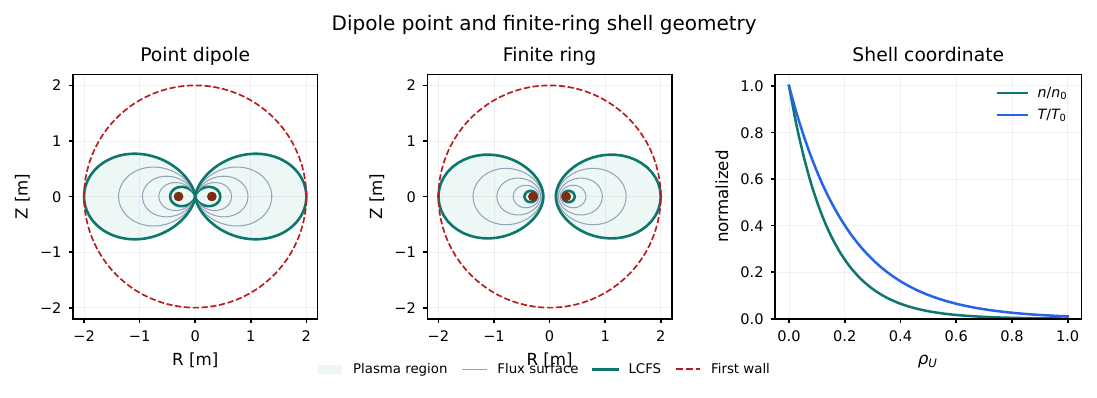}
\caption{Dipole geometry modes. The point-dipole and finite-ring panels show
source-generated flux shells, pale cyan plasma regions inside the outer
closed shells, and a spherical outer-wall proxy. The profile
panel shows the shell-volume coordinate used by the marginal-stability
profile.}
\label{fig:dipole-geometry}
\end{figure}

The dipole branch integrates over flux shells. It supports both a point
dipole proxy and a finite-current-loop vacuum-field correction, shown in
Fig.~\ref{fig:dipole-geometry}. For the point-dipole model, a field line
crossing the equator at \(L\) follows
\begin{equation}
  r = L\cos^2\lambda ,
\end{equation}
and the volume enclosed by shells out to \(L\) is analytic:
\begin{equation}
  V_{\rm enc}(L)=\frac{64\pi}{105}L^3,
  \qquad
  \frac{\dd V}{\dd L}=\frac{64\pi}{35}L^2 .
\end{equation}
The finite-ring mode replaces the point-dipole flux with the exact current
loop flux contour through the equatorial point. In dimensionless coordinates
\(\lambda=L/r_{\rm ring}\), the source tabulates
\begin{equation}
  \psi(\rho,z)=
  \frac{\sqrt{\rho}}{k}\left[(2-k^2)K(k)-2E(k)\right],
  \qquad
  k^2=\frac{4\rho}{(1+\rho)^2+z^2},
\end{equation}
and derives \(B(\lambda)\), \(V(\lambda)\), and the flux-tube specific volume
\(U=\left|\dd V/\dd\psi\right|\). The physical calibration preserves the
input field meaning through
\begin{equation}
  \mu_0 I = 4 r_{\rm ring}B_{\rm ring}.
\end{equation}
The finite-ring volume is obtained by sorting axisymmetric grid cells by
their vacuum flux value and accumulating \(2\pi R\,\Delta R\,\Delta Z\).
The specific volume then follows from the exact axisymmetric identity
\begin{equation}
  U=\oint\frac{\dd l}{B}=\left|\frac{\dd V}{\dd\psi}\right|,
\end{equation}
so the 0-D calculation does not need to trace each field line
profile integral.

The profile model is tied to dipole interchange physics rather than to a
minor-radius exponent. The flux-tube volume scales as \(U(L)\sim L^4\) for
the point-dipole limit. A marginal adiabatic profile with
\(\delta(pU^\gamma)=0\), \(\gamma=5/3\), gives
\begin{equation}
  p(L)\propto L^{-20/3}.
\end{equation}
The implemented density and temperature split is
\begin{equation}
  n(L)\propto U^{-1}, \qquad
  T(L)\propto U^{-2/3}.
\end{equation}
The normalized shell coordinate used for profile integration is
\begin{equation}
  \rho_U =
  \frac{\ln U-\ln U_{\rm in}}
       {\ln U_{\rm out}-\ln U_{\rm in}},
\end{equation}
with beta evaluated on inner and outer shells as
\(\beta(L)=2\mu_0 p(L)/B(L)^2\). The dipole branch reports inner and outer
beta, \(U\)-ratio, shell-integrated radiation, and \(n_0\tauE\). It does not
implement a predictive \(\tauE\) closure. Consequently, the dipole POPCON
should be read as a power-balance map conditional on the supplied confinement
time and profile assumptions, consistent with the levitated-dipole motivation
of Hasegawa and related work \cite{hasegawa1987}.
Two deliberately simple proxies are exposed in the result fields. The wall
area is the spherical proxy \(\Sw=4\pi R_p^2\), used only to form wall-load
diagnostics. The cyclotron model is labeled
\texttt{equatorial\_shell\_proxy}: it weights shell emission by the equatorial
field \(B(L)^{2.5}\) rather than integrating \(B^{2.5}\) along the full field
line. These approximations are part of the dipole branch's trust boundary.

\subsection{Stellarator}

\begin{figure*}[t]
\centering
\includegraphics[width=\textwidth]{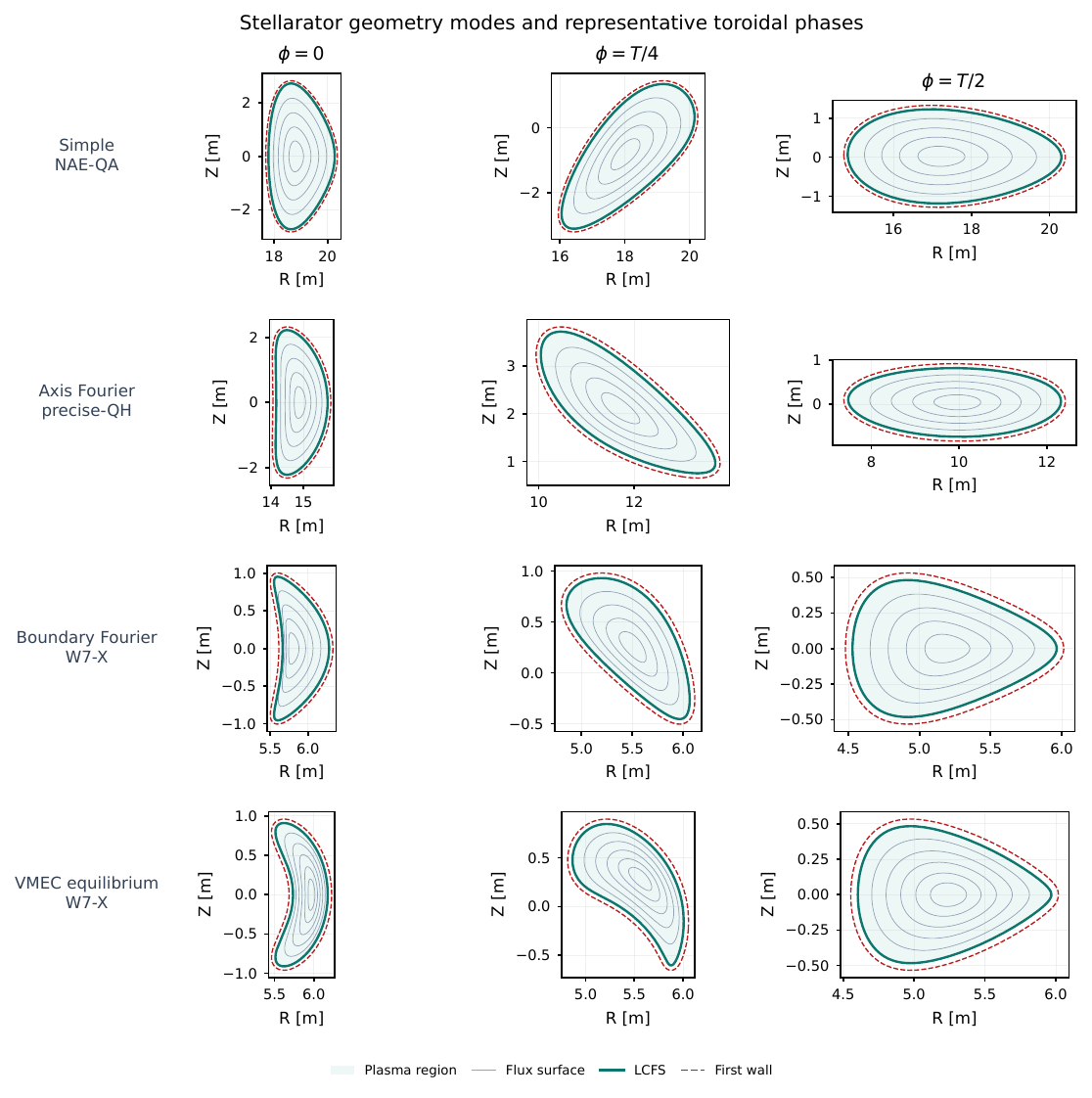}
\caption{Stellarator geometry routes. Rows show a simple near-axis concept
shape, the precise-QH near-axis construction, an explicit W7-X
boundary-Fourier model, and an imported W7-X equilibrium. Columns show three
toroidal phases for each route; pale cyan shading marks the plasma region
inside each teal boundary. The imported-equilibrium row uses nested
flux surfaces supplied by the source equilibrium reader rather than
synthetic display surfaces.}
\label{fig:stellarator-modes}
\end{figure*}

The stellarator branch combines four geometry routes: a simple near-axis
concept model, a first- and second-order near-axis construction, explicit
boundary-Fourier paths for machine-like shapes, and imported VMEC/DESC
equilibria when available
\cite{garren1991,landreman2019,landreman2021,hirshman1983,panici2022}.
Figure~\ref{fig:stellarator-modes} shows representative cross-sections for
each route. The device choices follow the source presets used in the
examples shown here: NAE-QA for the simple concept path, precise-QH for
the higher-order near-axis path, and W7-X for boundary and imported
equilibrium paths.

For the boundary path, the code integrates one field period and multiplies
by the number of field periods. A boundary function
\((R(\theta,\phi),Z(\theta,\phi))\) gives the plasma volume
\begin{equation}
  \Vp =
  N_{\rm fp}
  \int_0^{2\pi/N_{\rm fp}}
  \left|
  \oint \frac{1}{2}R^2\,\dd Z
  \right|\dd\phi ,
\end{equation}
with a corresponding surface-area integral over the boundary or wall offset.
Interior display/profile surfaces for boundary-Fourier shapes use the same
modes but fade the non-axis terms toward the magnetic axis,
\begin{align}
  R(\rho,\theta,\phi) &= R_0
  +a\sum_{m,n} f_m(\rho)R_{mn}\cos(m\theta-nN_{\rm fp}\phi),\\
  Z(\rho,\theta,\phi) &=
  a\sum_{m,n} f_m(\rho)Z_{mn}\sin(m\theta-nN_{\rm fp}\phi),
\end{align}
with \(f_0=1\), \(f_{\pm1}=\rho\), and
\(f_{|m|\ge2}=\rho^{1.5}\). This keeps the \(\rho=1\) boundary unchanged
while making high-order bean/triangular shaping round off toward the axis.
Near-axis concept geometry uses a Fourier magnetic axis,
\begin{equation}
  R_{\rm ax}(\varphi)=\sum_n R_n\cos(nN_{\rm fp}\varphi),
  \qquad
  Z_{\rm ax}(\varphi)=\sum_n Z_n\sin(nN_{\rm fp}\varphi),
\end{equation}
and a flux-conserving cross-section in the Frenet frame. The axis derivatives
used for curvature and torsion are evaluated analytically from the Fourier
series; in cylindrical basis the calculation uses
\begin{equation}
\bm d_1=(R_0',R_0,Z_0'),\quad
\bm d_2=(R_0''-R_0,2R_0',Z_0''),\quad
\bm d_3=(R_0'''-3R_0',3R_0''-R_0,Z_0'''),
\end{equation}
with
\begin{equation}
  \frac{\dd\ell}{\dd\varphi}
  =\sqrt{R_0^2+R_0'^2+Z_0'^2},\quad
  \kappa=\left\|\frac{\dd\hat t}{\dd\ell}\right\|,\quad
  \tau=\frac{\bm d_1\cdot(\bm d_2\times\bm d_3)}
  {\|\bm d_1\times\bm d_2\|^2}.
\end{equation}
The higher-order path solves the periodic near-axis sigma equation,
\begin{equation}
  \frac{\dd\sigma}{\dd\varphi}
  +(\iota+N_{\rm hel}N_{\rm fp})
  \left(\frac{\bar{\eta}^4}{\kappa^4}+1+\sigma^2\right)
  -2\frac{\bar{\eta}^2}{\kappa^2}
  (-\tau)\frac{G_0}{B_0}=0,
\end{equation}
with the on-axis rotational transform \(\iota\) solved as an unknown. Boundary
models use Fourier surfaces of the form
\begin{align}
  R(\theta,\phi) &= R_{00}
  +a\sum_{m,n}R_{mn}\cos(m\theta-nN_{\rm fp}\phi),\\
  Z(\theta,\phi) &=
  a\sum_{m,n}Z_{mn}\sin(m\theta-nN_{\rm fp}\phi).
\end{align}
Imported-equilibrium paths use VMEC or DESC boundary coefficients, nested
surfaces, volume, surface area, and available field-strength moments so that
the 0-D power account is anchored to a known geometry rather than to a
low-order analytic proxy. For imported equilibria the source reads surfaces
and field harmonics in the form
\begin{align}
  R(s,\theta,\phi) &=
  \sum_{m,n}R_{mn}(s)\cos(m\theta-nN_{\rm fp}\phi)
  +\sum_{m,n}\tilde R_{mn}(s)\sin(m\theta-nN_{\rm fp}\phi),\\
  Z(s,\theta,\phi) &=
  \sum_{m,n}Z_{mn}(s)\sin(m\theta-nN_{\rm fp}\phi)
  +\sum_{m,n}\tilde Z_{mn}(s)\cos(m\theta-nN_{\rm fp}\phi),\\
  |B|(s,\theta,\phi) &=
  \sum_{m,n}B^c_{mn}(s)\cos(m\theta-nN_{\rm fp}\phi)
  +\sum_{m,n}B^s_{mn}(s)\sin(m\theta-nN_{\rm fp}\phi).
\end{align}
When the Jacobian harmonics are present, the nonuniform-field factor is
computed as the flux-volume average
\begin{equation}
  {\cal G}_{B,p}=
  \frac{\int (|B|/B_0)^p |\sqrt g|\,\dd s\,\dd\theta\,\dd\phi}
       {\int |\sqrt g|\,\dd s\,\dd\theta\,\dd\phi}.
\end{equation}

The first-order near-axis cross-section is written in Frenet normal/binormal
coordinates as
\begin{equation}
  \Delta\bm P_1 =
  X_{1c}\cos\theta\,\hat n+
  \left(Y_{1s}\sin\theta+Y_{1c}\cos\theta\right)\hat b,
\end{equation}
with
\begin{equation}
  X_{1c}=\bar\eta/\kappa,\qquad
  Y_{1s}=\kappa/\bar\eta,\qquad
  Y_{1c}=\kappa\sigma/\bar\eta.
\end{equation}
The optional second-order ``bean'' correction adds
\(\rho^{1.5}\Delta\bm P_2\) with
\begin{equation}
  \Delta\bm P_2 =
  \left(X_{20}+X_{2c}\cos2\theta+X_{2s}\sin2\theta\right)\hat n+
  \left(Y_{20}+Y_{2c}\cos2\theta+Y_{2s}\sin2\theta\right)\hat b,
\end{equation}
plus a display cap, so high-order shaping
appears near the edge without distorting the magnetic-axis core. If the
second-order construction fails its internal checks, the branch falls back to
the first-order ellipse and marks the geometry path in the outputs.

The same boundary integrator supplies \(\Vp\), plasma surface area, and wall
area. The wall offset uses the signed shoelace orientation of each poloidal
section to choose the outward normal, rather than a centroid-to-vertex rule.
This is important for W7-X-like bean or crescent sections whose concave
notches would make a centroid rule point inward locally. Internal profile
weights are built by integrating scaled interior surfaces, enforcing monotone
volume, and defining
\begin{equation}
  \rho=\sqrt{V_{\rm enclosed}/\Vp},\qquad
  \dd V = 2\Vp\rho\,\dd\rho.
\end{equation}
Thus the familiar \(1/(1+S)\) profile averages are a consequence of the
volume-radius definition, not an assumption of circular cross-section.

The stellarator branch exposes ISS04 confinement diagnostics and Sudo
density-margin checks \cite{yamada2005,sudo1990}. These are used as
operating-limit indicators in the POPCON layer. The implemented formulas are
\begin{align}
  \tau_{\rm ISS04} &=
  0.134\,f_{\rm ren}\,
  a_{\rm vol}^{2.28}R_0^{0.64}P_L^{-0.61}
  (\bar n/10^{19})^{0.54}B_0^{0.84}\iota_{2/3}^{0.41},\\
  n_{\rm Sudo} &=
  0.25\left(\frac{P_LB_0}{a_{\rm vol}^2R_0}\right)^{1/2}
  10^{20}\ {\rm m^{-3}},
\end{align}
with \(H_{\rm ISS04}=\tauE/\tau_{\rm ISS04}\). Here \(\iota_{2/3}\) denotes
the rotational transform at normalized radius \(2/3\); VSC uses the supplied
measured value or the near-axis value as the effective transform for this
zero-dimensional closure. The branch reports both the
empirical line average \(\bar n\) used by ISS04/Sudo and the geometry-derived
line average obtained from the actual volume-radius table. These diagnostics
are not a substitute for neoclassical transport, alpha confinement,
island/divertor physics, coil optimization, or full 3-D MHD equilibrium
analysis.

\subsection{Cross-Configuration Power-Account Differences}

The five branches share Eq.~\eqref{eq:pheat}, but not the same profile
weights, magnetic-field moments, or transport closure. Table~\ref{tab:power-side-by-side}
summarizes the model-level distinctions that are easiest to miss
when only the final POPCON panel is viewed.

\begin{table}[t]
\centering
\small
\caption{Configuration-specific paths inside the common 0-D power account.}
\label{tab:power-side-by-side}
\begin{tabularx}{\linewidth}{@{}p{0.16\linewidth}XXX@{}}
\toprule
Branch & Fusion source & Cyclotron field treatment & Transport closure \\
\midrule
Tokamak &
Closed-torus \(\rho=\sqrt{V/\Vp}\) integral with \(n_1n_2\sv(T_i)\). &
Uniform Trubnikov estimate times Miller or G-EQDSK \(B^{2.5}\) moment. &
\(E_{\rm th}/\tauE\), or \(\tauE\) solved from IPB98, ST, or ITPA20. \\
Mirror &
Self-similar cylindrical profile over center cell and throats. &
Dimensional axial \(\int B^{2.5}\dd V\), emphasizing high-field throats. &
Manual \(\tauE\), or assembled end loss with \(\phi_i+\phi_e+T_i+T_e\) over \(\tau_m\). \\
FRC &
Rigid-rotor \(G_2n_m^2\) source with density fixed by magnetic pressure. &
Independent \(G_{B25}\) moment of the reversed-field profile. &
Manual \(\tauE\), or LSX-style empirical \(\tauE\). \\
Dipole &
Flux-shell integral over \(L\) or finite-ring \(\lambda\), with \(U\)-based profiles. &
Equatorial shell proxy \(B(L)^{2.5}\). &
Manual \(\tauE\) only. \\
Stellarator &
Volume-radius integral from near-axis, boundary, or imported equilibrium geometry. &
Near-axis or equilibrium \(B^{2.5}\) moment when enabled. &
\(E_{\rm th}/\tauE\), or \(\tauE\) solved from ISS04. \\
\bottomrule
\end{tabularx}
\end{table}

\FloatBarrier
\section{POPCON, Interface, and Performance}
\label{sec:popcon}

\begin{figure*}[t]
\centering
\includegraphics[width=\textwidth]{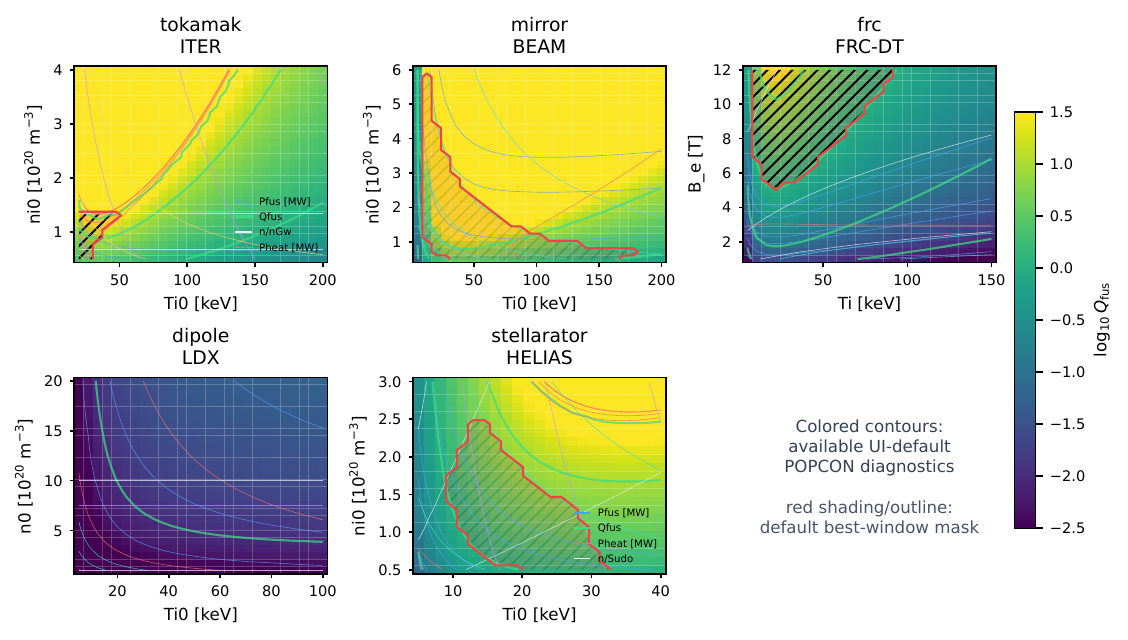}
\caption{Five POPCON-style scans regenerated from the stated physical inputs. The
background color is \(\log_{10}\qfus\). Colored contours are the default
diagnostic quantities shown by the application front end, including
fusion power, heating power, gain, density, beta, or configuration-specific
limits where defined. Red shading and outlines show the default best-window
mask for each configuration. \textbf{The shaded regions are conditional
operating-window masks under the stated default assumptions; they are not
feasibility predictions.} The panels are not equally predictive
(Table~\ref{tab:config-maturity}): in particular the dipole panel must be read
as a map over the \emph{supplied} confinement time $\tauE$, because the dipole
branch implements no predictive confinement closure.}
\label{fig:popcon}
\end{figure*}

The scan engine varies two named inputs and evaluates each grid point
independently. Validation failures, domain failures, and non-finite solver
outputs are stored as invalid values. The default plotting quantity is
\(\log_{10}\qfus\). Figure~\ref{fig:popcon} uses each configuration's
default scan axes, displayed physical quantities, and
default operating-window thresholds. The window mask has the form
\begin{equation}
  {\cal W}_{ij}
  =
  {\cal V}_{ij}
  \prod_k {\bf 1}\!\left[g_k(\bm{x}_{ij})\ {\rm satisfies}\ c_k\right],
\end{equation}
where \({\cal V}_{ij}\) is the valid-solver-output mask and \(g_k,c_k\) are
configuration-owned diagnostics and thresholds. The shaded regions in
Fig.~\ref{fig:popcon} use the same default best-window conditions exposed by
the front end; Table~\ref{tab:popcon-windows} lists them explicitly.

\begin{table*}[t]
\centering
\caption{Default POPCON best-window criteria used for the shaded regions in Fig.~\ref{fig:popcon}. All rows also require a valid solver output; invalid or non-finite points are excluded before thresholds are applied.}
\label{tab:popcon-windows}
\footnotesize
\setlength{\tabcolsep}{4pt}
\begin{tabular}{@{}p{0.13\textwidth}p{0.50\textwidth}p{0.28\textwidth}@{}}
\toprule
Configuration & Default shaded criteria & Additional available criteria, off by default \\
\midrule
tokamak & valid output; $P_{\rm fus}$ [MW] $\ge$ 10; $Q_{\rm fus}$ $\ge$ 1; $\bar{n}/n_{\rm GW}$ $\le$ 1; $P_{\rm heat}$ [MW] $\le$ 100 & $q_{95}$ $\ge$ 3; $\beta_N$ $\le$ 3.5; $H_{98}$ $\le$ 1.5 \\
mirror & valid output; $P_{\rm fus}$ [MW] $\ge$ 1; $Q_{\rm fus}$ $\ge$ 1; $\beta$ $\le$ 0.6 & none available \\
frc & valid output; $P_{\rm fus}$ [MW] $\ge$ 1; $Q_{\rm fus}$ $\ge$ 1; $\bar{s}$ $\ge$ 2 & $s/E$ $\le$ 4 \\
dipole & valid output; $P_{\rm fus}$ [MW] $\ge$ 1; $Q_{\rm fus}$ $\ge$ 1; $\beta_{\rm in}$ $\le$ 1 & none available \\
stellarator & valid output; $P_{\rm fus}$ [MW] $\ge$ 10; $Q_{\rm fus}$ $\ge$ 1; $\bar{n}/n_{\rm Sudo}$ $\le$ 1; $\beta_T$ $\le$ 0.05; $H_{\rm ISS04}$ $\le$ 1.5 & none available \\
\bottomrule
\end{tabular}
\end{table*}
 
The colored contours are display diagnostics, not feasibility claims. The
shaded windows should be interpreted conservatively. They encode only the
default thresholds in Table~\ref{tab:popcon-windows}: gain and
fusion power for every branch, plus density, heating, beta, Sudo, ISS04, or
FRC kinetic-parameter constraints where the branch defines them. Some
additional available redlines are off by default, such as tokamak \(q_{95}\),
\(\beta_N\), \(H_{98}\), and FRC \(s/E\). Other default-off contour families
remain display diagnostics only. None of these masks include every piece of
physics that could invalidate a point. A useful reading order is therefore:
first check whether the point is valid; then inspect \(\pheat\), \(\tauE\),
and the confinement-quality diagnostics; then apply the branch redlines;
finally identify which omitted physics is most likely to move or erase the
apparent window.

\paragraph{How to read the five panels.}
The tokamak panel is the closest to a conventional POPCON map because its
window can be conditioned on established empirical confinement scalings,
density limits, beta-like diagnostics, and optional imported equilibria.
Even there, the panel is a core-plasma operating-space map: pedestal, divertor,
current-drive, and plant-integration constraints are outside the mask.

The mirror panel is dominated by the selected end-loss model, beta limit,
mirror ratio, and alpha-deposition assumption. A favorable region should be
checked against the Pastukhov/gas-dynamic regime flag, the end-loading
diagnostics, and the alpha-deposition sensitivity described above; otherwise
the same contour can move from self-heated to externally heated.

The FRC panel should be read through the rigid-rotor closure. Density is tied
to magnetic pressure, the LSX-style \(\tauE\) is an empirical estimate, and
the kinetic parameter \(\bar{s}\) or \(s/E\) constrains only part of the stability
story. The shaded region is therefore a compact-toroid power-account window,
not a demonstration of tilt-stable or sustainment-compatible operation.

The dipole panel is the most parametric panel. It maps the supplied
confinement time, marginal profile, shell volume, wall proxy, and equatorial
cyclotron proxy. Because no predictive \(\tauE\) closure is implemented, the
dipole window answers ``what if this confinement time were achieved?'' rather
than ``where will the device confine?''

The stellarator panel combines ISS04 confinement and Sudo density indicators with the chosen
geometry source. Near-axis and VMEC/DESC geometry routes make the shape and
field moments more inspectable, but the mask still omits neoclassical
transport, island/divertor physics, alpha confinement, and coil feasibility.
Its role is to expose which assumptions matter before higher-fidelity
stellarator tools are invoked.

\subsection{Web Interface and Runtime}
\label{sec:web-runtime}

\begin{figure*}[t]
\centering
\includegraphics[width=\textwidth]{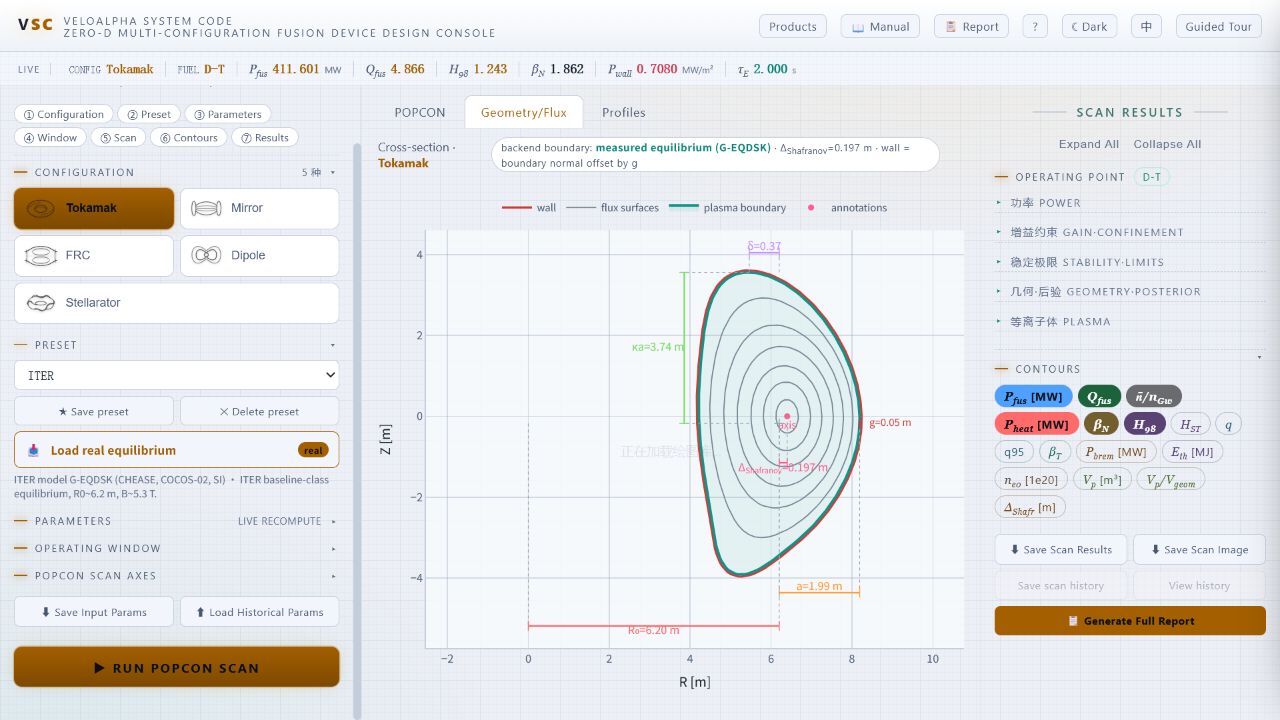}
\caption{VSC browser interface at
\url{https://hub.veloalpha.cn/vsc/}, shown in English mode. The ITER preset is
displayed in the Geometry/Flux view after loading the supplied G-EQDSK
equilibrium; the panel displays the
magnetic axis, flux surfaces, LCFS, first-wall offset, and geometric
annotations returned by the live back end.}
\label{fig:platform-homepage}
\end{figure*}

\begin{figure*}[t]
\centering
\includegraphics[width=\textwidth]{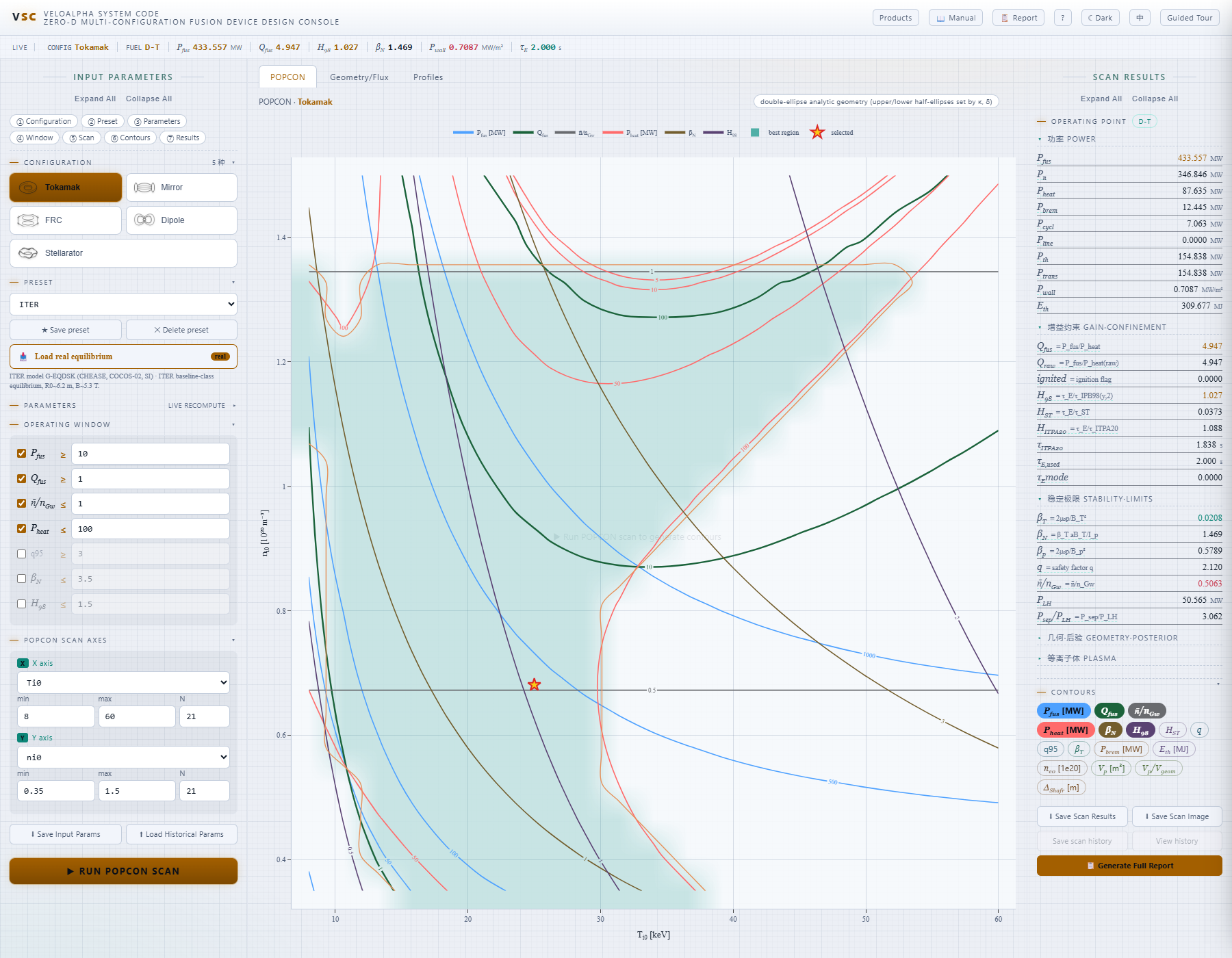}
\caption{ITER POPCON view generated with the same imported G-EQDSK geometry
as Fig.~\ref{fig:platform-homepage}. The scan covers
$T_{i0}=8$--$60$~keV and $n_{i0}=0.35$--$1.50\times10^{20}$~m$^{-3}$ on a
$21\times21$ grid. The left panel shows the ITER preset, operating-window
criteria, and scan axes; the right panel expands the power,
gain--confinement, and stability groups. Secondary parameter and derived
geometry/plasma groups are collapsed to keep the principal labels readable.
The star marks the selected ITER operating point.}
\label{fig:platform-popcon-expanded}
\end{figure*}

The input and output quantities used by the manuscript calculations are also
available through a browser interface. V1 is available at
\url{https://hub.veloalpha.cn/vsc/}. Figures~\ref{fig:platform-homepage}
and~\ref{fig:platform-popcon-expanded} show the ITER geometry and the
corresponding POPCON view. The page provides the
configuration selector, preset selector, editable parameters, operating-window
controls, POPCON scan launch, geometry/profile tabs, saved scan history, and
report-generation entry point around the same \texttt{run\_case} and
\texttt{scan2d} solver interface used for the figures in this work.

Table~\ref{tab:runtime} gives representative warm-run timings on the local
Windows workstation used for this manuscript update (Intel Core Ultra 9
275HX, Python 3.12.13). A \texttt{run\_case} call includes the display geometry
payload, whereas \texttt{scan2d} evaluates the scalar solver across the grid
without regenerating that full payload at every point. This distinction
explains why geometry-heavy dipole and stellarator single-point calls are
slower while their POPCON grids remain interactive.

\begin{table}[t]
\centering
\caption{Representative VSC runtime after one warmup call. Single-point
entries are medians of seven calls; scan entries are medians of three
51-by-51 grids (2601 points). Values are hardware- and environment-specific
user expectations, not physics-validation metrics. Repetition ranges and the
source hash are recorded in \texttt{data/runtime\_benchmark\_20260713.json}.}
\label{tab:runtime}
\footnotesize
\setlength{\tabcolsep}{5pt}
\begin{tabular}{@{}lrr@{}}
\toprule
Preset & \texttt{run\_case} & 51-by-51 \texttt{scan2d} \\
\midrule
Tokamak ITER & 0.73 ms & 1.96 s \\
Mirror BEAM & 0.40 ms & 0.92 s \\
FRC-DT & 0.86 ms & 0.58 s \\
Dipole LDX & 266 ms & 1.47 s \\
Stellarator W7-X & 75 ms & 0.68 s \\
\bottomrule
\end{tabular}
\end{table}

\section{Verification and Application Scope}
\label{sec:verification}

\subsection{Kernel and Geometry Verification}

Verification claims are scoped to individual quantities. Table~\ref{tab:external-anchors}
lists selected published, analytic, and independent-code anchors:
Bosch--Hale reactivities to $\le 3.2\%$ at the sampled temperatures, the IPB98(y,2) confinement time at the
ITER inductive point to $\sim 1\%$, the Greenwald limit, the Pastukhov barrier
factor, and the near-axis rotational transform, curvature, and torsion
cross-checked against pyQSC at $10^{-13}$--$10^{-14}$. These comparisons support
only the listed kernel or geometry quantity; they do not establish experimental
validation of a complete power-balance pathway.

The pyQSC comparison, for example, supports the stellarator \emph{geometry}
sub-module only. It does not cover the ISS04 closure, Sudo limit, radiation
integrals, or profile weights. Likewise, analytic-versus-numeric agreement
checks an implemented identity rather than validating it against an experiment.
None of these anchors is presented as a synchronized cross-code POPCON benchmark.

\begin{table*}[t]
\centering
\caption{Selected verification anchors for implemented physics kernels and
geometry modules. \textbf{Type} uses lit. for a VSC result compared with a
published value or a cited formula evaluated at the stated parameters, code
for independent-code comparisons, and ana. for analytic internal-consistency
checks. A lit. row is not necessarily a number quoted verbatim from a paper;
the Reference column states when the cited formula has been independently
evaluated for this work.}
\label{tab:external-anchors}
\footnotesize
\setlength{\tabcolsep}{4pt}
\begin{tabularx}{\textwidth}{@{}p{0.12\textwidth}Xp{0.13\textwidth}p{0.16\textwidth}p{0.10\textwidth}p{0.06\textwidth}@{}}
\toprule
Module & Benchmark quantity & VSC & Reference & Agreement & Type \\
\midrule
Reactivity &
D--T $\langle\sigma v\rangle$ at $10$~keV &
$1.136\times10^{-22}$ &
$1.13\times10^{-22}$~m$^3$/s \cite{bosch1992} &
$0.5\%$ & lit. \\
Reactivity &
maximum relative deviation from the Bosch--Hale fit at $T=5,10,20,50,100$~keV
(D--T and D--$^3$He) and $T=5,10,20,50$~keV (D--D) &
$3.2\%$ &
Bosch--Hale \cite{bosch1992} &
max. $3.2\%$ & lit. \\
Tokamak &
IPB98(y,2) $\tau_E$, ITER inductive &
$3.62$~s &
$\sim 3.66$~s \cite{iter1999} &
$\sim 1.1\%$ & lit. \\
Tokamak &
Greenwald density limit, ITER &
$1.194\times10^{20}$~m$^{-3}$ &
$1.194\times10^{20}$~m$^{-3}$ from $I_p/(\pi a^2)$ at the ITER inputs
\cite{greenwald1988} &
same to shown digits & lit. \\
Mirror &
Pastukhov $G(R_m{=}100)$ vs $\ln(4R_m{+}1)$ &
$6.026$ &
$5.994$ from the large-$R_m$ asymptote \cite{pastukhov1974} &
$0.53\%$ & ana. \\
Stellarator &
near-axis $\iota$, Landreman--Sengupta \S5.1 &
$0.4183069102$ &
$0.4183069102$ (pyQSC) \cite{landreman2019} &
$10^{-13}$ & code \\
Stellarator &
axis curvature and torsion along the magnetic axis &
pointwise data &
pyQSC data \cite{landreman2019} &
max. diff. $10^{-14}$ & code \\
FRC &
rigid-rotor moments, analytic vs numeric &
match &
own quadrature &
$<10^{-5}$ & ana. \\
FRC &
average-$\beta$ theorem $\tanh K/K = 1-x_s^2/2$ &
match &
closed form \cite{tuszewski1988} &
exact & ana. \\
Dipole &
shell volume, analytic vs numeric &
$9.40778$ &
$9.40778$~m$^3$ &
$<10^{-6}$ & ana. \\
Dipole &
pressure profile for marginal interchange stability &
$p(L)\propto L^{-20/3}$ &
analytic profile \cite{hasegawa1987} &
exact & ana. \\
\bottomrule
\end{tabularx}
\end{table*}
 
The figures and tables are generated from the same solver interfaces described
in Sec.~\ref{sec:arch}, so the plotted quantities follow the documented power
account and configuration-specific closures.

\paragraph{Scope-level context.}

Table~\ref{tab:popcon-comparison} compares purpose and scope rather than
claiming numerical equivalence. cfspopcon and OpenPOPCON are closer to
traditional tokamak POPCON workflows \cite{cfspopcon,openpopcon}. PROCESS
occupies a broader plant-systems design role \cite{processcode}, while FUSE
provides an integrated Julia framework spanning plasma, engineering, control,
and plant-design actors \cite{meneghini2024fuse}. VSC
instead uses a common calculation framework: it keeps a single scan layer
and makes five different 0-D closures visible through one interface. This
tradeoff is deliberate. VSC prioritizes cross-configuration transparency,
while specialized tokamak tools retain the depth required for detailed
scenario studies.

\begin{table*}[t]
\centering
\caption{Scope-level comparison with existing POPCON or systems-code workflows. The table is feature based; no cross-code numerical benchmark is claimed here.}
\label{tab:popcon-comparison}
\footnotesize
\setlength{\tabcolsep}{3pt}
\begin{tabular}{@{}p{0.15\textwidth}p{0.16\textwidth}p{0.23\textwidth}p{0.34\textwidth}@{}}
\toprule
Tool & Main scope & Strength & Contrast with VSC \\
\midrule
cfspopcon \cite{cfspopcon} & Tokamak POPCON analysis & Rich tokamak-oriented physics workflow and notebooks & VSC trades tokamak-specific depth for a common power-balance treatment of five magnetic configurations and direct geometry views. \\
OpenPOPCON \cite{openpopcon} & Compact open tokamak POPCON model & Transparent single-route tokamak operating contours & VSC keeps the POPCON approach but extends it to mirror, FRC, dipole, and stellarator models. \\
PROCESS \cite{processcode} & Tokamak and stellarator power-plant systems optimization & Broad plasma and engineering plant models and constraints & VSC covers five magnetic topologies through a common reduced power-balance and scan interface rather than a full plant optimizer. \\
FUSE \cite{meneghini2024fuse} & Integrated fusion pilot-plant design & Coupled plasma, engineering, control, and plant-design actors & VSC retains a reduced 0-D scope and uses the ITER-named cases only for a qualitative operating-space comparison. \\
\bottomrule
\end{tabular}
\end{table*}
 
\subsection{FUSE ITER Operating-Space Comparison}
\label{sec:external-popcon}

The external graphical comparison is deliberately limited to ITER-named
cases. The panel places a locally executed FUSE v1.1.3 scalar-initialization
sweep (Julia 1.11.5) beside the native VSC ITER scan over the same plotted
temperature--density rectangle \cite{meneghini2024fuse}. Other candidate
panels are omitted because the available device cases and scan coordinates
cannot be mapped closely enough for an interpretable comparison; the
selection is not based on whether their contour shapes agree with VSC.

This restriction is important because a POPCON plot is not device-agnostic.
Changing the geometry, density definition, impurity assumptions, confinement
closure, alpha deposition model, radiation model, or gain denominator can
move the apparent favorable region. In addition, this FUSE sweep initializes
temperature and pedestal density while retaining the ITER actuator set at a
fixed total launched auxiliary power of 87.4~MW. Its plotted ratio is
\(Q_{\rm FUSE,plot}=P_{\rm fus,total}/P_{\rm aux,launched}\), whereas VSC
solves the required external heating at each point and plots
\(Q_{\rm VSC}=P_{\rm fus}/P_{\rm heat}\). The figure therefore provides a
qualitative comparison of operating-space trends for two ITER-named reduced
calculations, not a conventional matched POPCON benchmark or pointwise
numerical validation.

\begin{figure*}[t]
\centering
\includegraphics[width=\textwidth]{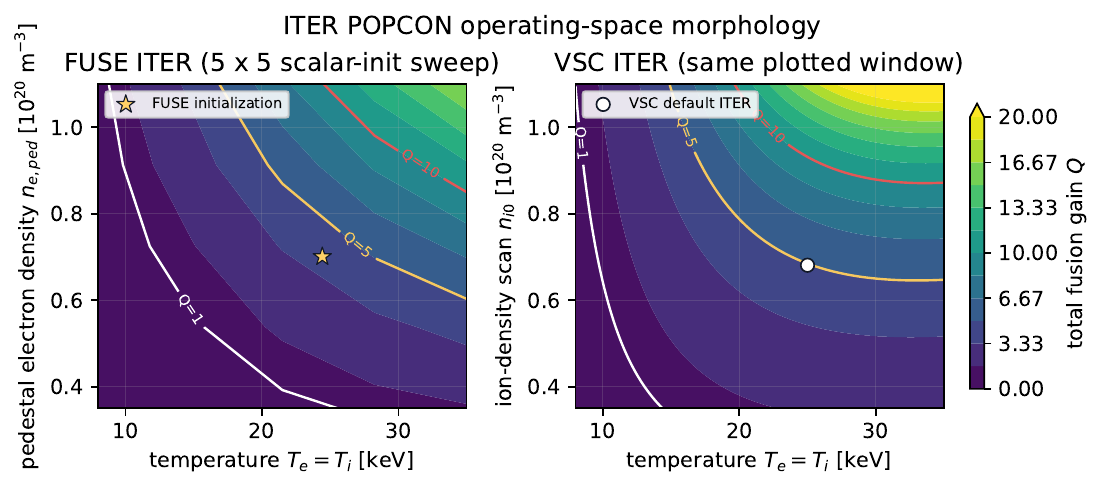}
\caption{ITER operating-space scans from FUSE (left) and VSC (right) over the
same numerical window, \(T_e=T_i=8\)--35~keV and density
\(0.35\)--\(1.10\times10^{20}\,\mathrm{m}^{-3}\). The FUSE ordinate is
pedestal electron density and its displayed ratio uses the fixed
\(P_{\rm aux,launched}=87.4\)~MW actuator input. The VSC ordinate is its ion
density scan coordinate; its default ITER point gives the independently
calculated requirement \(P_{\rm heat}=87.63\)~MW and
\(Q_{\rm fus}=4.95\). These values come from the local runs rather than from
Ref.~\cite{meneghini2024fuse}. The broadly similar contour trend is
qualitative evidence only because geometry, profiles, composition, density
coordinates, and gain denominators are not matched. The star in the FUSE
panel marks its scalar-initialization point, while the white circle in the
VSC panel marks the independently evaluated VSC default ITER point.}
\label{fig:external-popcon-fuse-iter}
\end{figure*}
\FloatBarrier
 
\subsection{Limitations and Trust Boundary}
\label{sec:limits}

VSC should be read as a transparent 0-D analysis and teaching framework.
It does not model blankets, tritium breeding, recirculating plant power,
materials, neutron damage, coil stress, startup, disruption mitigation,
divertor solutions, fast-particle distributions, non-Maxwellian effects,
pellet or fueling dynamics, control, or full 3-D equilibrium fidelity. Its
non-tokamak branches are especially assumption-sensitive because empirical
confinement databases are smaller, geometry is less standardized, and some
closures are best read as bounding or pedagogical estimates.

The practical trust boundary is:
\begin{enumerate}
  \item The code can compare 0-D trends under declared inputs and geometry
  weights.
  \item The code can expose invalid points, redline violations, and
  sensitivity to \(\tauE\), density, temperature, and field assumptions.
  \item The code cannot certify a reactor concept or replace
  higher-fidelity transport, stability, equilibrium, alpha, engineering,
  or economics calculations.
\end{enumerate}
Several display or proxy choices are intentionally marked in outputs rather
than hidden. FRC interior nested surfaces are midplane-flux-anchored display
curves and do not enter the power account. Dipole wall area and cyclotron
loss use the spherical-wall and equatorial-shell proxies described above.
Non-imported stellarator interior surfaces are low-order near-axis or
boundary-fade constructions; imported equilibria carry real nested surfaces
only to the extent supplied and parsed by the equilibrium reader. These
distinctions are important because a visually smooth geometry panel is not
the same thing as a high-fidelity equilibrium or transport calculation.
Table~\ref{tab:branch-omissions} lists the branch-specific omissions that are
most likely to move or remove a candidate operating window.

\begin{table*}[t]
\centering
\caption{Branch-specific omissions that can move or remove a candidate POPCON operating window.}
\label{tab:branch-omissions}
\footnotesize
\setlength{\tabcolsep}{4pt}
\begin{tabularx}{\textwidth}{@{}p{0.12\textwidth}XX@{}}
\toprule
Branch & Omitted or proxy physics & Consequence for interpretation \\
\midrule
Tokamak &
Pedestal and ELM physics; current drive and bootstrap current; divertor and impurity-exhaust constraints; startup and disruption limits &
Tokamak windows are the most mature in the present framework, but remain plasma-core operating windows rather than integrated reactor scenarios. \\
Mirror &
Sheath and secondary-electron physics; nonclassical radial transport; detailed alpha slowing-down loss-cone scattering; end-expander engineering &
End-loss and alpha-deposition assumptions can change required heating and end loading even when the 0-D gain contour is favorable. \\
FRC &
Tilt-mode nonlinear evolution; rotational instability dynamics; flux-tube end shortening; full kinetic-MHD stability; sustainment/current-drive model &
The LSX-style confinement estimate and rigid-rotor equilibrium support reduced-order concept studies, not stability-certified FRC operation. \\
Dipole &
Predictive \(\tauE\) closure; finite-orbit-width effects; non-spherical wall geometry; full field-line cyclotron integration &
Dipole POPCON panels map supplied confinement time and proxy geometry assumptions; they should not be read as predictions. \\
Stellarator &
Neoclassical transport; island/divertor physics; alpha confinement; coil optimization; applicability limits of ISS04 to optimized quasi-symmetry &
Near-axis and VMEC geometry anchors are useful, but the 0-D pathway is not a full stellarator transport calculation. \\
\bottomrule
\end{tabularx}
\end{table*}
 \FloatBarrier

Within that boundary, VSC is useful precisely because it refuses to
hide the assumptions. A POPCON panel is treated as a map of conditional
calculations, not as a verdict.

\section{Summary and Outlook}

This work establishes VSC as a common zero-dimensional platform for examining
fusion power balance across tokamaks, magnetic mirrors, field-reversed
configurations, dipoles, and stellarators. The central result is not that these
configurations share the same confinement physics, but that their different
geometry, profile, radiation, transport, and operating-limit assumptions can
be expressed within one traceable calculation sequence. A common power account
therefore remains comparable across configurations, while each branch retains
its own native radial coordinate, magnetic-field moment, confinement closure,
and interpretation boundary. This separation is essential for using a unified
platform without implying uniform physical maturity.

The present calculations demonstrate three practical capabilities. First,
single-point solutions expose the complete balance among fusion power,
charged-particle deposition, transport loss, bremsstrahlung, impurity
radiation, cyclotron or synchrotron loss, and external heating. Second, POPCON
scans propagate the same calculation over density--temperature or other
configuration-appropriate planes while isolating invalid points and applying
explicit operating-window conditions. Third, geometry models ranging from
analytic boundaries to imported G-EQDSK and VMEC/DESC equilibria connect the
0-D balance to inspectable volume, surface, profile, and magnetic-field
weights. The selected checks against published formulas, reaction-rate fits,
analytic identities, and pyQSC golden data support the individual quantities
listed in Sec.~\ref{sec:verification}. The ITER comparison with FUSE shows
that both reduced scans produce rising gain with increasing temperature and
density and broadly similar low-to-intermediate-gain contour morphology. It
does not establish agreement of heating power: FUSE holds its launched
auxiliary power fixed, whereas VSC calculates the external heating required
by its own power balance.

These results define where VSC is most useful in an early design study. It can
rapidly identify the combinations of density, temperature, field, geometry,
and assumed confinement that dominate a candidate operating region; it can
also reveal when an attractive region depends on a prescribed confinement
time, a proxy radiation treatment, or an incomplete stability condition. The
tokamak branch currently has the strongest predictive basis, whereas the
mirror and stellarator branches rely on reduced empirical closures, the FRC
branch uses a partial empirical scaling, and the dipole branch remains
parametric in \(\tau_E\). VSC should therefore be used to organize assumptions,
compare trends, and select cases for deeper analysis, not to certify a complete
reactor design.

The present version is a first step toward a more complete fusion-device design
environment. Near-term work will broaden experimental and cross-code
validation, quantify uncertainty and sensitivity, and strengthen the
configuration-specific closures. Subsequent development will couple the 0-D
power balance to higher-dimensional equilibrium, transport, stability,
heating, fast-particle, and engineering models so that promising points can be
passed systematically to higher-fidelity calculations. A documented
application programming interface (API), namely a machine-readable route for
exchanging data with VSC, will allow users and external programs to submit,
edit, or upload input parameter sets and retrieve structured results. The
longer-term roadmap also includes AI-assisted interpretation and a
natural-language agent for constructing studies and examining assumptions.
Together, these developments will preserve the speed and transparency of the
0-D layer while extending VSC toward increasingly realistic and integrated
fusion-device design.

\section*{Acknowledgments}
The authors are grateful to Dr. Tailin Wu and Dr. Caoxiang Zhu for their
insightful discussions and constructive suggestions.

\bibliographystyle{unsrt}

\end{document}